\documentclass[aps,prl,twocolumn,superscriptaddress]{revtex4-1}
\usepackage{graphicx}
\usepackage{dcolumn}
\usepackage{bm}
\usepackage{amsmath}
\usepackage{verbatim}
\usepackage{amssymb}
\usepackage{listings}
\usepackage{color}
\usepackage{epstopdf}
\renewcommand\vec[1]{\boldsymbol{#1}}
\newcommand\Fig[1]{Fig.~\ref{#1}}

\newcommand\pt[1]{\left({#1}\right)}

\let\ifincludesupplements\iftrue
\let\ifnotbuildingseparatesupp\iftrue

\begin{document}

\title{Lattice glass model in three spatial dimensions}
\author{Yoshihiko Nishikawa}
\email[]{yoshihiko.nishikawa@umontpellier.fr}
\affiliation{Laboratoire Charles Coulomb, UMR 5221 CNRS, Universit\'e de Montpellier, 
34095 Montpellier, France}
\affiliation{Department of Basic Science, Graduate School of Arts and Sciences, \\
The University of Tokyo, 3-8-1 Komaba, Meguro, Tokyo 153-8902, Japan}
\author{Koji Hukushima}
\email[]{k-hukushima@g.ecc.u-tokyo.ac.jp}
\affiliation{Department of Basic Science, Graduate School of Arts and Sciences, \\
The University of Tokyo, 3-8-1 Komaba, Meguro, Tokyo 153-8902, Japan}
\affiliation{
Komaba Institute for Science, The University of Tokyo, 3-8-1 Komaba, Meguro, Tokyo 153-8902, Japan}

\date{\today}

\begin{abstract}
The understanding of thermodynamic glass transition has been hindered by the lack of proper 
models beyond mean-field theories. 
Here, we propose a three-dimensional lattice glass model on a simple cubic lattice that 
exhibits the typical dynamics observed in fragile supercooled liquids such as two-step 
relaxation, super-Arrhenius growth in the relaxation time, and dynamical heterogeneity.
Using advanced Monte Carlo methods, we compute the thermodynamic properties  
deep inside the glassy temperature regime, well below the onset temperature of the slow 
dynamics. The specific heat has a finite jump towards the thermodynamic limit with critical 
exponents close to those expected from the hyperscaling and the random first-order transition 
theory for the glass transition. We also study an effective free energy of glasses, the 
Franz--Parisi potential, as a function of the overlap between equilibrium and quenched 
configurations. The effective free energy indicates the existence of a first-order phase 
transition, consistent with the random first-order 
transition theory. These findings strongly suggest that the glassy dynamics of the model has its 
origin in thermodynamics.
\end{abstract}


\maketitle

A thermodynamic (or ideal) glass transition in finite low dimensions has 
been actively discussed both theoretically and experimentally for decades since the 
seminal works by Kauzmann \cite{Kauzmann1948}, and Adam and Gibbs \cite{Adam1965}. 
A mean-field theory, the random first-order transition (RFOT) theory of structural 
glasses, proposed and developed in 
Refs.~\cite{Kirkpatrick1987,Kirkpatrick1987a,Kirkpatrick1988,Kirkpatrick1989} 
indeed shows that a thermodynamic glass transition at finite temperature exists 
with vanishing configuration entropy, or complexity 
\cite{Berthier2011}. In finite dimensions, numerical 
simulation of models of fragile supercooled liquids is a promising way to theoretically 
explore a possibility of the thermodynamic glass transition. However, the notoriously long 
relaxation time prevents us to access directly low-temperature thermodynamics of 
glass-forming supercooled liquids. Although recent progress on particle models 
simulated using the swap Monte Carlo method \cite{Berthier2016,Ninarello2017} 
allows us to get much more stable glass configurations at lower temperature or higher 
density, the thermodynamic glass transition is still inaccessible.

In an analogy to phase transitions into long-range ferromagnetic and crystal states, the 
lower critical dimension for the thermodynamic glass transition in models with discrete 
symmetry may be lower than that in models with continuous symmetry. Thus, 
it would be crucial to explore a possibility of a thermodynamic glass transition in 
finite-dimensional lattice models with discrete symmetry. Although several simple 
lattice models with the mean-field thermodynamic glass transition have already been 
proposed \cite{Biroli2001,Ciamarra2003,McCullagh2005}, they are not fully suitable to 
study the finite-dimensional glass transition: For the models in 
Refs.~\cite{Biroli2001,McCullagh2005}, their mean-field glass transitions are turned into a 
crossover in finite dimensions or their low-temperature glassy states are unstable due 
to crystallization. The other model \cite{Ciamarra2003}, while having the typical glassy 
dynamics, is a monodisperse model, where an efficient 
algorithm such as the swap method \cite{Berthier2016,Ninarello2017} 
is not known. Another lattice glass model was proposed in Ref.~\cite{Sasa2012}, which 
is shown to have irregular configurations as an ordered state. However, its autocorrelation 
function decays without any plateau even at low temperature. Thus the model does not 
have the essential features of the structural glasses.

To find finite-dimensional models that allow us 
to access equilibrium low-temperature states without crystallization is still actively discussed.
In this letter, we propose a simple lattice glass model to study glassy behaviors in three 
dimensions. Our model, in contrast to several lattice models 
\cite{Biroli2001,McCullagh2005,Sasa2012}, shows typical two-step relaxation dynamics 
as observed in fragile supercooled liquids. We study a binary mixture of the model, 
where the non-local swap dynamics explained below provide the benefits for equilibration.
By large-scale Monte Carlo simulations, we equilibrate the system at temperature well below 
that where two-step relaxation emerges. We study the effect of a coupling field conjugate to 
the overlap between the system and a quenched configuration using the Wang--Landau 
algorithm \cite{Wang2001,Wang2001a}, and compute the quenched version of the Franz--Parisi 
potential \cite{Franz1995,Franz1997,Franz1998}, an effective 
free energy of the glass transition. Our results show that  the system indeed has a
thermodynamic behavior consistent with the RFOT theory.

\begin{figure}[tbp]
\center \includegraphics[width=\linewidth]{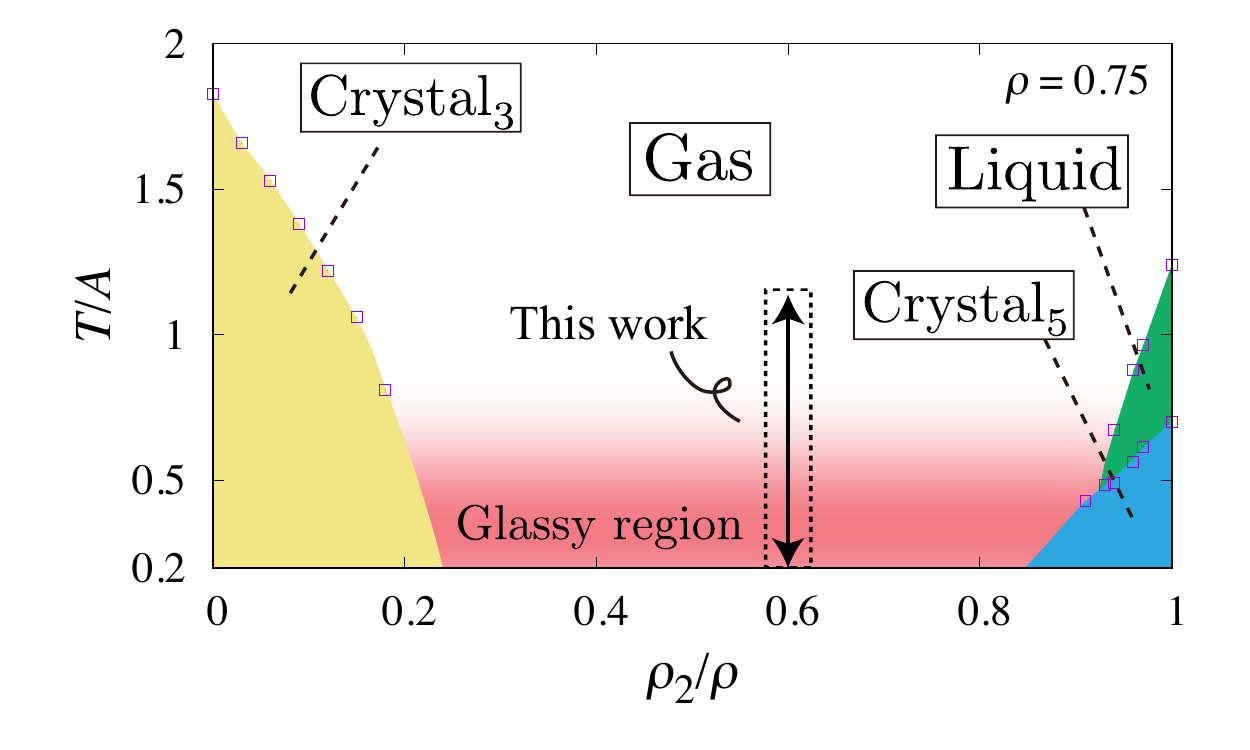}
\caption{
Phase diagram of the model in the plane of temperature $T$ and type-2
particle density $\rho_2$ with the total density $\rho=0.75$.
Crystal$_3$ and Crystal$_5$ stand crystalline phases
where a crystal structure is mainly composed of particles with 
$\ell_1 = 3$ and $\ell_2 = 5$, respectively. 
The gas, liquid, crystalline phases are separated by first-order phase transitions. In the 
glassy region, the system shows an aging effect and its autocorrelation function 
relaxes in two steps. We study the system with $\rho_2 / \rho = 0.6$ in this 
work. 
}
\label{fig:PD}
\end{figure}

A lattice glass model we study in this letter
is a binary mixture of particles defined by the Hamiltonian with a positive constant $A$
\begin{equation}
H \pt{\vec n, \vec \sigma} = A\sum_{i=1}^{L^3} n_i
\Big(\sum_{j\in \partial i}n_j - \ell_{\sigma_i}\Big)^2,
\label{eq:Ham}
\end{equation}
where the first summation runs over all the lattice sites. The lattice is the simple cubic 
lattice with linear dimension $L$. We denote the occupancy of 
site $i$ with $n_i \in \{ 0, 1\}$, and the type of a particle at site $i$ with $\sigma_i\in\{1,2\}$. 
The parameter $\ell_{\sigma_i}$ determines the most favorable number of neighboring 
particles of type $\sigma_i$ particle. Here, we consider a binary mixture of $\ell_1 = 3$ 
and $\ell_2 = 5$ particles. The boundary conditions in all the 
directions are periodic. We study this model at finite temperature $T$ by Monte Carlo 
simulation that preserves the number of each type of particles \cite{Kawasaki1966}:  
Two randomly chosen particles are 
swapped or a randomly chosen particle is moved to a vacant site of the lattice 
chosen also randomly with the Metropolis probability. Depending on the density of each 
type of particle, $\rho_1$ and $\rho_2$, with fixed total density $\rho = \rho_1+ \rho_2 = 0.75$, 
our model has crystal phases at low temperature, see \Fig{fig:PD}. 
When $\rho_1 = 0.3$ and $\rho_2 = 0.45$, we confirmed the absence of a 
drop in the energy and a large peak in the specific heat in the glassy region, indicating no 
crystallization. 

Our model is somewhat related to a softened model of a lattice glass
proposed by Biroli and M\'ezard (BM) \cite{Biroli2001,Foini2011}.
The Hamiltonian of their model $H_\mathrm{BM} = A\sum_{i=1} 
n_i \theta\pt{\sum_{j\in \partial i}n_j  - \ell_{\sigma_i}} \pt{\sum_{j\in \partial i}n_j  - \ell_{\sigma_i}}$
with the Heaviside step function $\theta(x)$.
The crucial difference between the two models is the number of neighboring 
particles that each particle energetically favors: In the soft BM model, any number of 
neighboring particles lower than $\ell_\sigma$ achieves the lowest energy 
while only $\ell_\sigma$ does in our model. This slight change makes the 
entropy of our model at low temperature smaller than that of the BM model
and the dynamics more similar to fragile supercooled liquids. 

The dynamics of our model is studied by a local Monte Carlo algorithm \cite{Kawasaki1966}, 
which usually gives dynamics qualitatively similar to molecular and Brownian dynamics. 
In our simulation, particles can move to only neighboring sites. 
While any physical quantity of our model relaxes rapidly at high temperature, the 
relaxation gets extremely slow with decreasing temperature. To quantify the slow 
dynamics of the model, we study the autocorrelation function 
\begin{equation}
C\pt{t; t_\mathrm w} = 
\Big\langle\frac{\frac1N\sum_{i} 
\delta_{\sigma_i \pt{t_\mathrm w}, \sigma_i \pt{t_\mathrm w + t}} - C_0}
{1 - C_0}\Big\rangle,
\label{eq:autocorrelation}
\end{equation}
where $t_\mathrm w$ is the waiting time, $C_0 = \rho (\rho_1^2 + \rho_2^2)$, $N =  \rho L^3$ the total number of particles,
and the summation is taken over sites with $n_i(t_\mathrm w)=1$.
We also measure the dynamical susceptibility $\chi_4 \pt{t; t_\mathrm w}$ characterizing the 
dynamical heterogeneity observed in supercooled liquids. In the limit $t_\mathrm w \to \infty$, 
the system is in equilibrium and we denote $C\pt{t} = C\pt{t; t_\mathrm w \to \infty}$ and 
$\chi_4\pt{t} = \chi_4\pt{t; t_\mathrm w \to \infty}$. We set $t_\mathrm w$ to a sufficiently 
large value to study the equilibrium dynamics of the model so that $C\pt{t; t_\mathrm w}$ 
and $C\pt{t; t_\mathrm w / 10}$ agree with each other. Typical values of $t_\mathrm w$ 
range from $10^2$ to $10^8$ Monte Carlo sweeps depending on temperature. 
Whereas the autocorrelation function of the system decays rapidly at high temperature, 
a two-step relaxation emerges in $C(t)$ at temperature lower than $T / A \simeq 0.6$, 
see \Fig{fig:AC} (see also \cite{Supplement} for a physical interpretation of the plateau in $C(t)$). 
The dynamical susceptibility $\chi_4(t)$ shows a peak at finite time, 
indicating the emergence of heterogeneous dynamics of the system 
\cite{Kob1997,Yamamoto1998,Ediger2000,Toninelli2005,Berthier2011}.
The peak value grows with decreasing temperature (see inset of \Fig{fig:chi4}),
and it suggests that the dynamics gets more heterogeneous at lower temperature.

The RFOT theory of the glass transition predicts 
slow dynamics frozen at a dynamical transition temperature without any thermodynamic 
anomaly. In the vicinity of the dynamical transition temperature, the relaxation time diverges
algebraically. However, an exponentially growing relaxation time well described by the 
Vogel--Fulcher--Tammann (VFT) law has been observed in 
experiments of fragile super-cooled liquids. Activated process in finite dimensions is supposed 
to wipe out the mean-field dynamical transition.
Here, we study temperature dependence of the relaxation time $\tau_\alpha$ measured as 
a time when $C\pt{t}$ decays to $e^{-1}$.
Around temperature where the two-step relaxation 
emerges, the relaxation time shows a super-Arrhenius growth with decreasing 
temperature (see \Fig{fig:reltime}).
We find that the VFT law fits our data at lower temperatures very well
with $T_\mathrm{VFT} / A = 0.177(6)$ (see also \cite{Supplement} for 
estimation of the dynamical transition temperature). Note that, as the relaxation time 
at low temperature increases with the system size, $T_\mathrm{VFT}$ would shift to higher 
temperature in larger systems \cite{Supplement}.

\begin{figure}[tbp]
\center \includegraphics[width=.9\linewidth]{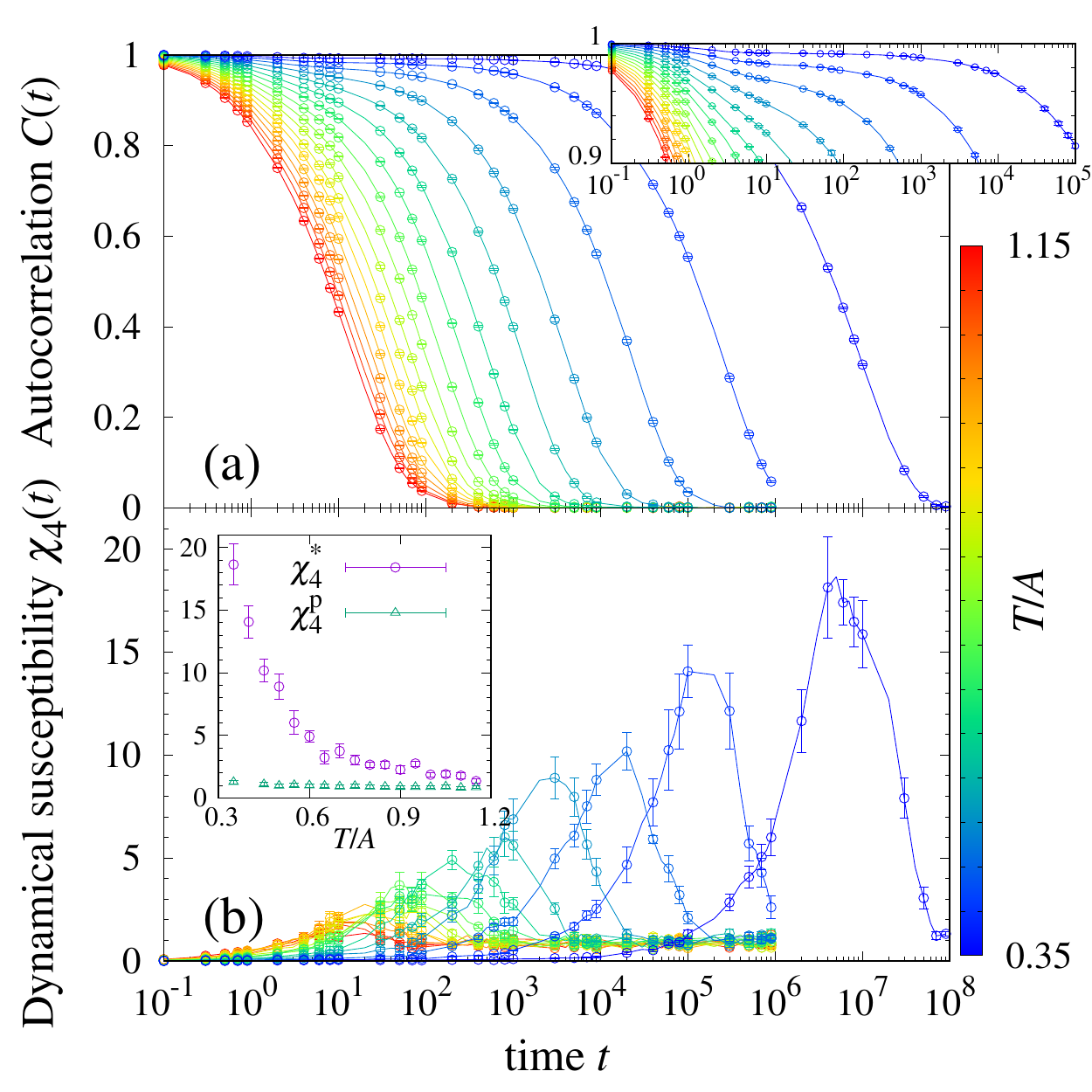}
\caption{
(a) The autocorrelation function of the system.
Inset shows an enlarged view around the plateau region. 
(b) The dynamical susceptibility $\chi_4 \pt{t}$ of the system.
Inset shows temperature dependence of the peak and the plateau values of $\chi_4\pt{t}$ 
that are denoted as $\chi_4^*$ and $\chi_4^\mathrm p$, respectively. 
The system size $N = 6000$ ($L = 20$).
}
\label{fig:AC}
\label{fig:chi4}

\center \includegraphics[width=\linewidth]{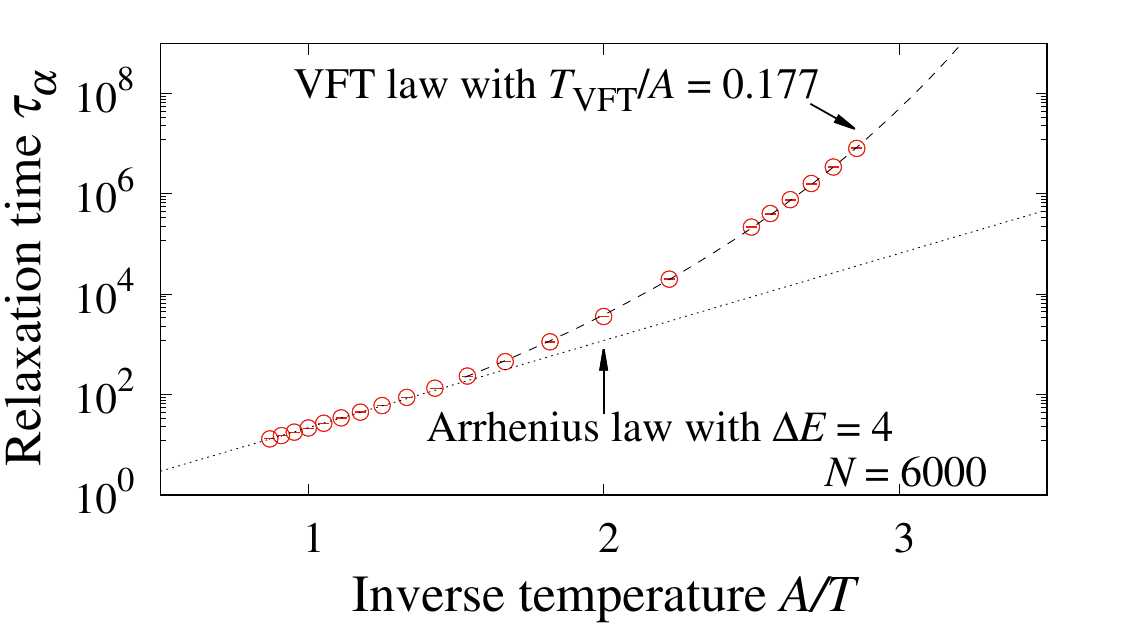}
\caption{
The Arrhenius plot of the relaxation time $\tau_\alpha$ of the system with $N = 6000$ ($L = 20$). 
The broken and dotted lines are the Arrhenius law with $\Delta E = 4$ and the VFT law with 
$T_{\rm VFT}/A=0.177$ respectively for guides to eyes.
}
\label{fig:reltime}
\end{figure}

\begin{figure}[bp]
\center \includegraphics[width=\linewidth]{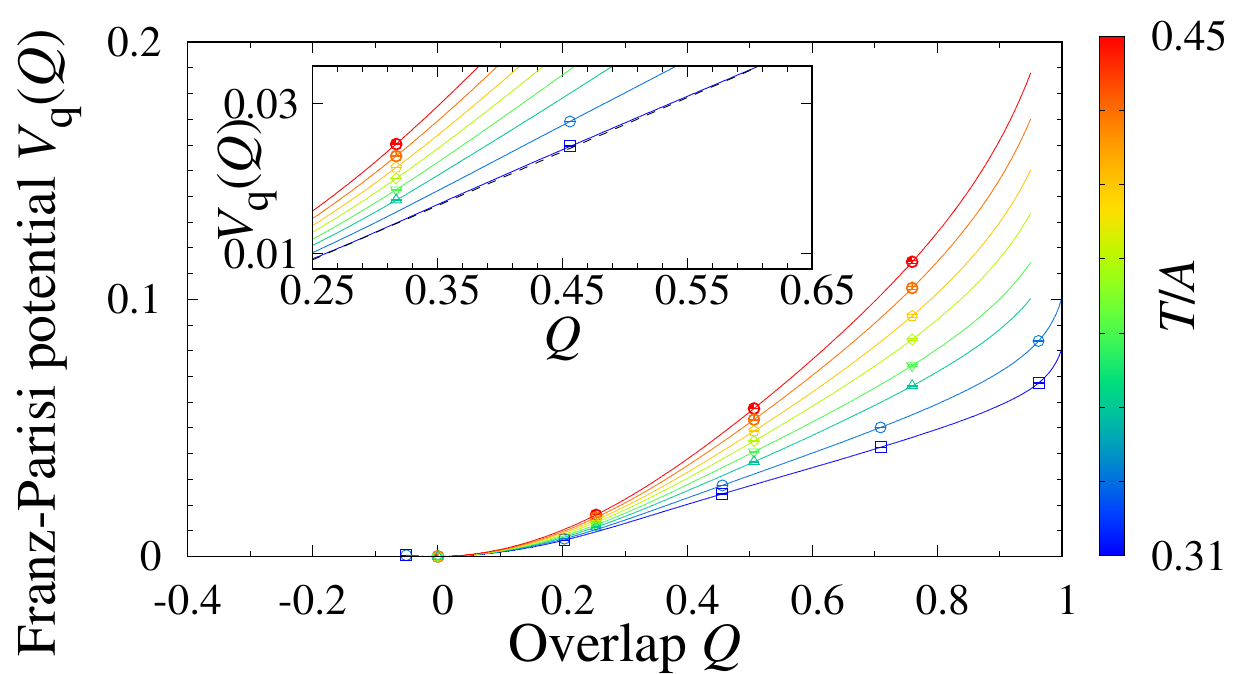}
\caption{
The Franz--Parisi potential $V_\mathrm q(Q)$ of the system 
with $N=1296$ ($L=12$). At high temperature
the potential is convex while it is slightly non-convex at low temperature, $T/A = 0.31$.
Inset shows an enlarged view. 
The broken line is a guide to eyes to show the non-convexity. 
}
\label{fig:FP}

\end{figure}

\begin{figure}[bp]
\includegraphics[width=\linewidth]{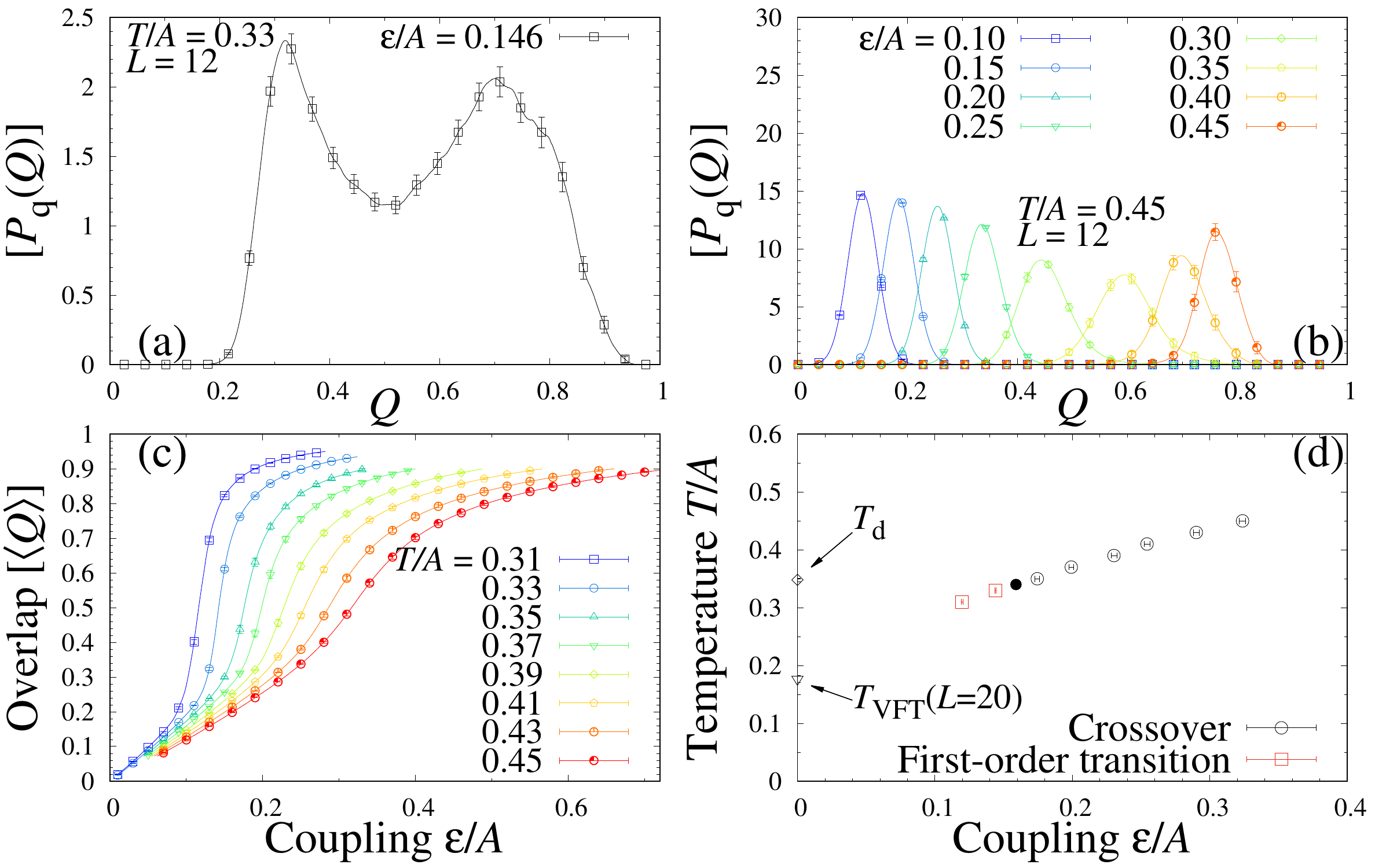}
\caption{
(a, b) The probability distribution of the overlap averaged over quenched reference 
configurations $[P_\mathrm q(Q)]$ with finite coupling field $\varepsilon$
at $T/A = 0.33$ (a) and $0.45$ (b), respectively. (c) The overlap $[\langle Q \rangle]$ 
as a function of the field $\varepsilon$. With increasing $\varepsilon$, $[\langle Q \rangle]$ 
grows smoothly at high temperature while it shows a rapid increase 
at low temperature. The system size $N=1296$ ($L=12$).
(d) Temperature-coupling phase diagram of the system. The empty points and their errors 
are estimated by the peak of the spin glass susceptibility $\chi_\mathrm{SG}$ from the 
data for $L=12$. The filled circle point represents an expected location of the critical point 
that is not determined precisely here.
The triangular point $T_\mathrm{VFT}$ is from the VFT-law fit of the relaxation time for 
$L=20$, see \Fig{fig:reltime}. The diamond point indicates the dynamical transition 
temperature $T_\mathrm c \simeq 0.348(4)$ estimated by fitting \cite{Supplement}.
}
\label{fig:epsQ}
\end{figure}

In the RFOT theory, the overlap between independent replicas has been studied as an 
order parameter for the thermodynamic glass transition. The overlap successfully detects the
glass transition in systems without spatial symmetry 
\cite{Cammarota2012,Cammarota2013,Kob2013,Ozawa2015}. The overlap can detect even 
the slow dynamics at low temperature 
through its effective free energy \cite{Franz1995,Franz1997,Franz1998}.
In our model, however, the overlap and its distribution function can have no 
anomaly at any temperature due to its spatial symmetry \cite{Mezard2012}.
The temperature dependence of the long-time limiting value of the dynamical susceptibility 
$\chi_4(t \to \infty)$, equivalent to the spin glass susceptibility 
$\chi_\mathrm{SG}$, is indeed almost independent of temperature (see inset of 
\Fig{fig:chi4}). This is in contrast to a Potts spin glass model \cite{Takahashi2015} 
where $\chi_{\rm SG}$ increases with approaching its transition temperature 
\cite{TakahashiHukushima}.

To study rare fluctuations and the effective free energy of the overlap appropriately, 
we introduce a field $\varepsilon$ in our model: A system at temperature $T$ is coupled 
by the field to a quenched configuration randomly sampled at the same temperature in 
equilibrium \cite{Franz1995,Franz1997,Franz1998}. 
The field explicitly breaks the invariance associated with the spatial symmetry, and thus the 
overlap can have nontrivial distributions. In the RFOT theory, the 
field induces a first-order transition at low temperature and the transition terminates at 
a critical point belonging to the random-field Ising model universality class \cite{Franz2013}. 
The Franz--Parisi potential reveals the existence of metastable states at
low-temperature and allows us to compute the configurational entropy 
as a free-energy difference between two minima \cite{Berthier2014,Berthier2017}. Here, we 
use the Wang--Landau algorithm \cite{Wang2001,Wang2001a} with the multi-overlap ensemble 
\cite{Berg1998} to compute the density of states $\Omega_T(Q)$ as a function of the 
overlap $Q$ at a given temperature $T$ with a fixed reference configuration. The 
Franz--Parisi potential $V_\mathrm q(Q)$ is computed from $\Omega_T(Q)$ 
by averaging over quenched reference configurations. We prepared reference configurations 
by equilibrating the system at each temperature for $10^8$ Monte Carlo sweeps with non-local 
swaps of particles. The non-local swap dynamics is faster with a factor of $\sim 10^2$ at 
low temperature than the local swap dynamics while the factor slightly depends on temperature 
and the system size. The number of reference configurations is $912$ for $T/A = 0.31$ and 
$0.33$, and $48$ for higher temperatures. At high temperature, the potential 
$V_\mathrm{q}(Q)$ is convex and the overlap $[\langle Q \rangle]$ as a function of the field 
$\varepsilon$ grows gradually, see \Fig{fig:FP} and \Fig{fig:epsQ} (c). Here, the brackets 
$\langle \cdot \rangle$ and $[\cdot]$ represent the thermal and the reference-configuration 
averages, respectively. At $T/A = 0.33$ and $0.31$, the Franz--Parisi 
potential is slightly non-convex (see \Fig{fig:FP}), and the probability distribution of the overlap 
$[P_\mathrm q(Q)]$ with finite $\varepsilon$ has two separated peaks whereas that at 
temperature higher than $T/A = 0.35$ shows a clear single peak at any $\varepsilon$, 
see \Fig{fig:epsQ} (a) and (b), respectively~\footnote{Note that the Franz--Parisi potential is convex at any temperature in the thermodynamic limit as the free-energy barrier is sub-extensive, and the non-convexity is seen only in finite systems.}.
We thus conclude that the coupling field 
induces a first-order transition into our model that terminates at a critical point at finite 
temperature as in mean-field and particle models in finite dimensions
\cite{Franz1997,Franz1998,Franz2013,Berthier2013,Berthier2015} (see \Fig{fig:epsQ} (d)
for $T$-$\varepsilon$ phase diagram).

When we assume the existence of the thermodynamic glass transition at 
$T_\mathrm K > 0$ for a system with spatial symmetry, we find the system-size 
dependence of the effective transition point $\varepsilon_c(L)$ and the 
configurational entropy density $s_\mathrm{conf}(L)$ as follows.
In the thermodynamic limit, an infinitesimal coupling field should make them 
have a finite overlap at $T < T_\mathrm K$ whereas the 
overlap is strictly zero when $\varepsilon = 0$ due to spatial symmetry \cite{Mezard2012}. 
In finite systems, the first-order transition at finite field thus never goes to the expected 
$T_\mathrm K$, even in the limit $\varepsilon \to +0$, while it 
may approach zero temperature in the limit. At temperature lower than 
$T_\mathrm K$, the effective transition point $\varepsilon_\mathrm c$ 
should scale as $\varepsilon_\mathrm c = O(L^{-a})$ while at higher temperature 
$\varepsilon_\mathrm c - \varepsilon_\mathrm c^{(\infty)} = O(L^{-a})$ where 
$\varepsilon_\mathrm c^{(\infty)} = \lim_{L\to \infty} \varepsilon_\mathrm{c}(L) > 0$ 
and $a > 0$, which depends on systems \cite{Fisher1982,Mueller2014}. Regarding 
that the configurational entropy density $s_\mathrm{conf}$ measured as the free 
energy difference in the 
Franz--Parisi potential is almost equivalent to $\varepsilon_\mathrm c$, we expect 
$s_\mathrm{conf}$ of finite systems to be finite even below expected $T_\mathrm K$, 
but decreases with $\sim L^{-a}$, as $\varepsilon_\mathrm c(L)$ does. Studying the 
finite-size dependence of the first-order transition line and the configurational entropy 
would be a decisive test for the RFOT theory.

\begin{figure}[tbp]
\center \includegraphics[width=\linewidth]
{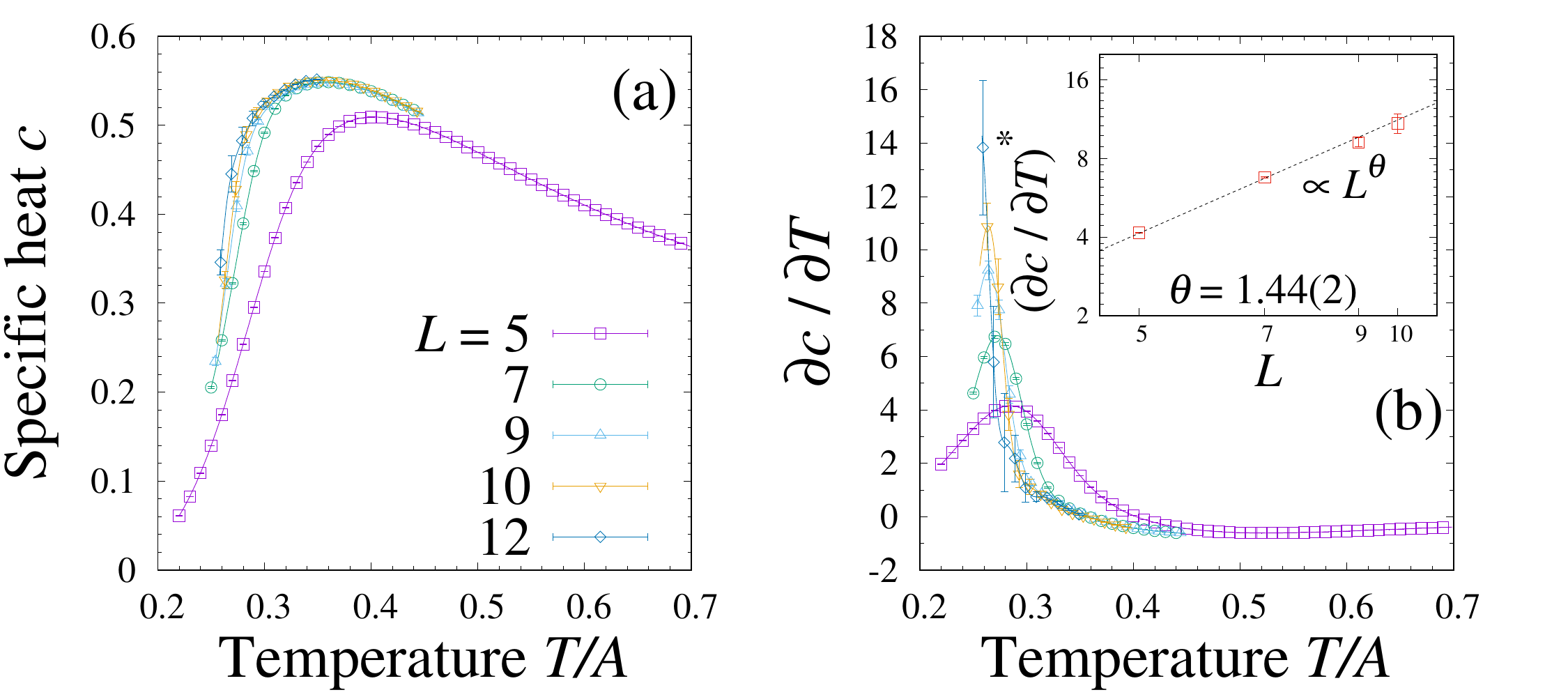}%
 \caption{
Temperature dependence of (a) the specific heat $c$ and 
(b) its temperature derivative $\partial c/\partial T$ of the system. 
Inset of (b) shows the peak value of $\partial c/\partial T$ as a function of $L$.
}
\label{fig:TD}
\end{figure}

To measure the thermodynamic properties at low temperature, we use the non-local swap
dynamics of randomly chosen pairs of particles and the exchange Monte Carlo (or parallel 
tempering) method \cite{Hukushima1996} to further enhance equilibration. We also utilize 
the multiple-temperature reweighting technique \cite{Ferrenberg1988,Ferrenberg1989,Munger1991}, 
which significantly improves the accuracy of Monte Carlo results. Typical number of Monte 
Carlo sweeps for equilibration range from $10^8$ to $10^{11}$ per site depending on the 
system size. 

At temperature $T / A \simeq 0.6$ where the two-step relaxation emerges, the 
specific heat is rather smooth, and it starts to drop without a divergent behavior 
at lower temperature (see \Fig{fig:TD} (a)). To study further the system-size 
dependence of the specific heat, we compute the temperature-derivative of the 
specific heat $\partial c / \partial T$. We find that $\partial c / \partial T$ grows 
with increasing the system size, indicating that the drop gets steeper with the 
system size. Similar size-dependence has been observed in a mean-field model 
with RFOT \cite{Billoire2005} 
and a three-dimensional Potts glass model that has a RFOT-like spin-glass transition
\cite{Takahashi2015}.

Assuming the hyperscaling relation of the critical exponents $d\nu = 2 - \alpha$
with $d$ the spatial dimension, we find that the peak value of $\partial c / \partial T$ 
follows the finite-size scaling relation
\begin{equation}
\pt{\frac{\partial c}{\partial T}}^* \propto L^{\theta}
\end{equation}
with $\theta = (\alpha + 1) / \nu$. We thus can evaluate the two conventional critical exponents 
from $\theta$ as $\nu = 3 / \pt{\theta + d}$ and $\alpha = 2 - d\nu = 2 - 3d / \pt{\theta + d}$. If 
the specific heat does not diverge but has a finite jump at a thermodynamic 
glass transition as in the RFOT theory, the critical exponent $\alpha = 0$, 
suggesting $\nu = 2 / d$. The peak value $\pt{\partial c / \partial T}^*$ thus grows algebraically 
with exponent $\theta = 1 / \nu = d / 2$. This provides a useful way for studying thermodynamic
anomaly with a finite-size scaling analysis even when an order parameter 
is unknown. Here, in our model, the exponent $\theta = 1.44(2)$ (see inset of \Fig{fig:TD} (b)),
implying $\nu = 0.68(1)$ and $\alpha = -0.03(1)$, marginally compatible with the RFOT theory.

In summary, we proposed a lattice glass model that is very stable against 
crystallization and shows the typical dynamics observed in fragile supercooled liquids 
at low temperature. Our numerical 
results show that a glass transition is detected by a singularity of the temperature-derivative 
of the specific heat with exponents compatible with the RFOT theory.
We also numerically computed the quenched version of the Franz--Parisi potential using the 
Wang--Landau algorithm. The potential shows very large overlap
fluctuations at low temperature, and a first-order transition was found in the coupled system that 
terminates at a critical point. The large overlap fluctuations strongly suggest that 
low-temperature glassy dynamics emerge from thermodynamics of the system, similar 
to the mean-field models with the RFOT. We thus conclude that our model is useful to study 
and further examine the mean-field RFOT predictions. Thanks to the lattice nature of 
our model, it is easy to study its mean-field solution using the cavity method and compare 
it with our results. To determine precisely the phase diagram 
both in mean-field and finite dimensions will constitute a crucial
test of the validity of the theory. 

\begin{acknowledgments}
The authors thank A.~Ikeda for many useful discussions. Y.N. is grateful to J.~Takahashi, 
T.~Takahashi, S.~Takabe, H.~Yoshino, M.~Ozawa, L.~Berthier and D.~Coslovich
for critical comments and useful discussions. This work was supported by JSPS KAKENHI 
Grant Numbers 17J10496, 17H02923 and 19H04125. Numerical simulation in this work has 
mainly been performed by using the facility of the Supercomputer Center, Institute for Solid 
State Physics, the University of Tokyo. 
\end{acknowledgments}

\bibliography{lattice_glass}

\providecommand{\noopsort}[1]{}\providecommand{\singleletter}[1]{#1}%
\begin{thebibliography}{49}%
\makeatletter
\providecommand \@ifxundefined [1]{%
 \@ifx{#1\undefined}
}%
\providecommand \@ifnum [1]{%
 \ifnum #1\expandafter \@firstoftwo
 \else \expandafter \@secondoftwo
 \fi
}%
\providecommand \@ifx [1]{%
 \ifx #1\expandafter \@firstoftwo
 \else \expandafter \@secondoftwo
 \fi
}%
\providecommand \natexlab [1]{#1}%
\providecommand \enquote  [1]{``#1''}%
\providecommand \bibnamefont  [1]{#1}%
\providecommand \bibfnamefont [1]{#1}%
\providecommand \citenamefont [1]{#1}%
\providecommand \href@noop [0]{\@secondoftwo}%
\providecommand \href [0]{\begingroup \@sanitize@url \@href}%
\providecommand \@href[1]{\@@startlink{#1}\@@href}%
\providecommand \@@href[1]{\endgroup#1\@@endlink}%
\providecommand \@sanitize@url [0]{\catcode `\\12\catcode `\$12\catcode
  `\&12\catcode `\#12\catcode `\^12\catcode `\_12\catcode `\%12\relax}%
\providecommand \@@startlink[1]{}%
\providecommand \@@endlink[0]{}%
\providecommand \url  [0]{\begingroup\@sanitize@url \@url }%
\providecommand \@url [1]{\endgroup\@href {#1}{\urlprefix }}%
\providecommand \urlprefix  [0]{URL }%
\providecommand \Eprint [0]{\href }%
\providecommand \doibase [0]{http://dx.doi.org/}%
\providecommand \selectlanguage [0]{\@gobble}%
\providecommand \bibinfo  [0]{\@secondoftwo}%
\providecommand \bibfield  [0]{\@secondoftwo}%
\providecommand \translation [1]{[#1]}%
\providecommand \BibitemOpen [0]{}%
\providecommand \bibitemStop [0]{}%
\providecommand \bibitemNoStop [0]{.\EOS\space}%
\providecommand \EOS [0]{\spacefactor3000\relax}%
\providecommand \BibitemShut  [1]{\csname bibitem#1\endcsname}%
\let\auto@bib@innerbib\@empty
\bibitem [{\citenamefont {Kauzmann}(1948)}]{Kauzmann1948}%
  \BibitemOpen
  \bibfield  {author} {\bibinfo {author} {\bibfnamefont {W.}~\bibnamefont
  {Kauzmann}},\ }\href@noop {} {\bibfield  {journal} {\bibinfo  {journal}
  {Chem. Rev.}\ }\textbf {\bibinfo {volume} {43}},\ \bibinfo {pages} {219}
  (\bibinfo {year} {1948})}\BibitemShut {NoStop}%
\bibitem [{\citenamefont {Adam}\ and\ \citenamefont {Gibbs}(1965)}]{Adam1965}%
  \BibitemOpen
  \bibfield  {author} {\bibinfo {author} {\bibfnamefont {G.}~\bibnamefont
  {Adam}}\ and\ \bibinfo {author} {\bibfnamefont {J.~H.}\ \bibnamefont
  {Gibbs}},\ }\href {\doibase 10.1063/1.1696442} {\bibfield  {journal}
  {\bibinfo  {journal} {The Journal of Chemical Physics}\ }\textbf {\bibinfo
  {volume} {43}},\ \bibinfo {pages} {139} (\bibinfo {year} {1965})}\BibitemShut
  {NoStop}%
\bibitem [{\citenamefont {Kirkpatrick}\ and\ \citenamefont
  {Thirumalai}(1987)}]{Kirkpatrick1987}%
  \BibitemOpen
  \bibfield  {author} {\bibinfo {author} {\bibfnamefont {T.~R.}\ \bibnamefont
  {Kirkpatrick}}\ and\ \bibinfo {author} {\bibfnamefont {D.}~\bibnamefont
  {Thirumalai}},\ }\href {\doibase 10.1103/PhysRevB.36.5388} {\bibfield
  {journal} {\bibinfo  {journal} {Physical Review B}\ }\textbf {\bibinfo
  {volume} {36}},\ \bibinfo {pages} {5388} (\bibinfo {year}
  {1987})}\BibitemShut {NoStop}%
\bibitem [{\citenamefont {Kirkpatrick}\ and\ \citenamefont
  {Wolynes}(1987)}]{Kirkpatrick1987a}%
  \BibitemOpen
  \bibfield  {author} {\bibinfo {author} {\bibfnamefont {T.~R.}\ \bibnamefont
  {Kirkpatrick}}\ and\ \bibinfo {author} {\bibfnamefont {P.~G.}\ \bibnamefont
  {Wolynes}},\ }\href {\doibase 10.1103/PhysRevB.36.8552} {\bibfield  {journal}
  {\bibinfo  {journal} {Physical Review B}\ }\textbf {\bibinfo {volume} {36}},\
  \bibinfo {pages} {8552} (\bibinfo {year} {1987})}\BibitemShut {NoStop}%
\bibitem [{\citenamefont {Kirkpatrick}\ and\ \citenamefont
  {Thirumalai}(1988)}]{Kirkpatrick1988}%
  \BibitemOpen
  \bibfield  {author} {\bibinfo {author} {\bibfnamefont {T.~R.}\ \bibnamefont
  {Kirkpatrick}}\ and\ \bibinfo {author} {\bibfnamefont {D.}~\bibnamefont
  {Thirumalai}},\ }\href {\doibase 10.1103/PhysRevA.37.4439} {\bibfield
  {journal} {\bibinfo  {journal} {Physical Review A}\ }\textbf {\bibinfo
  {volume} {37}},\ \bibinfo {pages} {4439} (\bibinfo {year}
  {1988})}\BibitemShut {NoStop}%
\bibitem [{\citenamefont {Kirkpatrick}\ \emph {et~al.}(1989)\citenamefont
  {Kirkpatrick}, \citenamefont {Thirumalai},\ and\ \citenamefont
  {Wolynes}}]{Kirkpatrick1989}%
  \BibitemOpen
  \bibfield  {author} {\bibinfo {author} {\bibfnamefont {T.~R.}\ \bibnamefont
  {Kirkpatrick}}, \bibinfo {author} {\bibfnamefont {D.}~\bibnamefont
  {Thirumalai}}, \ and\ \bibinfo {author} {\bibfnamefont {P.~G.}\ \bibnamefont
  {Wolynes}},\ }\href {\doibase 10.1103/PhysRevA.40.1045} {\bibfield  {journal}
  {\bibinfo  {journal} {Physical Review A}\ }\textbf {\bibinfo {volume} {40}},\
  \bibinfo {pages} {1045} (\bibinfo {year} {1989})}\BibitemShut {NoStop}%
\bibitem [{\citenamefont {Berthier}\ and\ \citenamefont
  {Biroli}(2011)}]{Berthier2011}%
  \BibitemOpen
  \bibfield  {author} {\bibinfo {author} {\bibfnamefont {L.}~\bibnamefont
  {Berthier}}\ and\ \bibinfo {author} {\bibfnamefont {G.}~\bibnamefont
  {Biroli}},\ }\href {\doibase 10.1103/RevModPhys.83.587} {\bibfield  {journal}
  {\bibinfo  {journal} {Reviews of Modern Physics}\ }\textbf {\bibinfo {volume}
  {83}},\ \bibinfo {pages} {587} (\bibinfo {year} {2011})}\BibitemShut
  {NoStop}%
\bibitem [{\citenamefont {Berthier}\ \emph {et~al.}(2016)\citenamefont
  {Berthier}, \citenamefont {Coslovich}, \citenamefont {Ninarello},\ and\
  \citenamefont {Ozawa}}]{Berthier2016}%
  \BibitemOpen
  \bibfield  {author} {\bibinfo {author} {\bibfnamefont {L.}~\bibnamefont
  {Berthier}}, \bibinfo {author} {\bibfnamefont {D.}~\bibnamefont {Coslovich}},
  \bibinfo {author} {\bibfnamefont {A.}~\bibnamefont {Ninarello}}, \ and\
  \bibinfo {author} {\bibfnamefont {M.}~\bibnamefont {Ozawa}},\ }\href
  {\doibase 10.1103/PhysRevLett.116.238002} {\bibfield  {journal} {\bibinfo
  {journal} {Physical Review Letters}\ }\textbf {\bibinfo {volume} {116}},\
  \bibinfo {pages} {238002} (\bibinfo {year} {2016})}\BibitemShut {NoStop}%
\bibitem [{\citenamefont {Ninarello}\ \emph {et~al.}(2017)\citenamefont
  {Ninarello}, \citenamefont {Berthier},\ and\ \citenamefont
  {Coslovich}}]{Ninarello2017}%
  \BibitemOpen
  \bibfield  {author} {\bibinfo {author} {\bibfnamefont {A.}~\bibnamefont
  {Ninarello}}, \bibinfo {author} {\bibfnamefont {L.}~\bibnamefont {Berthier}},
  \ and\ \bibinfo {author} {\bibfnamefont {D.}~\bibnamefont {Coslovich}},\
  }\href {\doibase 10.1103/PhysRevX.7.021039} {\bibfield  {journal} {\bibinfo
  {journal} {Physical Review X}\ }\textbf {\bibinfo {volume} {7}},\ \bibinfo
  {pages} {021039} (\bibinfo {year} {2017})}\BibitemShut {NoStop}%
\bibitem [{\citenamefont {Biroli}\ and\ \citenamefont
  {M{\'{e}}zard}(2001)}]{Biroli2001}%
  \BibitemOpen
  \bibfield  {author} {\bibinfo {author} {\bibfnamefont {G.}~\bibnamefont
  {Biroli}}\ and\ \bibinfo {author} {\bibfnamefont {M.}~\bibnamefont
  {M{\'{e}}zard}},\ }\href {\doibase 10.1103/PhysRevLett.88.025501} {\bibfield
  {journal} {\bibinfo  {journal} {Physical Review Letters}\ }\textbf {\bibinfo
  {volume} {88}},\ \bibinfo {pages} {025501} (\bibinfo {year}
  {2001})}\BibitemShut {NoStop}%
\bibitem [{\citenamefont {Ciamarra}\ \emph {et~al.}(2003)\citenamefont
  {Ciamarra}, \citenamefont {Tarzia}, \citenamefont {de~Candia},\ and\
  \citenamefont {Coniglio}}]{Ciamarra2003}%
  \BibitemOpen
  \bibfield  {author} {\bibinfo {author} {\bibfnamefont {M.~P.}\ \bibnamefont
  {Ciamarra}}, \bibinfo {author} {\bibfnamefont {M.}~\bibnamefont {Tarzia}},
  \bibinfo {author} {\bibfnamefont {A.}~\bibnamefont {de~Candia}}, \ and\
  \bibinfo {author} {\bibfnamefont {A.}~\bibnamefont {Coniglio}},\ }\href
  {\doibase 10.1103/PhysRevE.67.057105} {\bibfield  {journal} {\bibinfo
  {journal} {Physical Review E}\ }\textbf {\bibinfo {volume} {67}},\ \bibinfo
  {pages} {057105} (\bibinfo {year} {2003})}\BibitemShut {NoStop}%
\bibitem [{\citenamefont {McCullagh}\ \emph {et~al.}(2005)\citenamefont
  {McCullagh}, \citenamefont {Cellai}, \citenamefont {Lawlor},\ and\
  \citenamefont {Dawson}}]{McCullagh2005}%
  \BibitemOpen
  \bibfield  {author} {\bibinfo {author} {\bibfnamefont {G.~D.}\ \bibnamefont
  {McCullagh}}, \bibinfo {author} {\bibfnamefont {D.}~\bibnamefont {Cellai}},
  \bibinfo {author} {\bibfnamefont {A.}~\bibnamefont {Lawlor}}, \ and\ \bibinfo
  {author} {\bibfnamefont {K.~A.}\ \bibnamefont {Dawson}},\ }\href {\doibase
  10.1103/PhysRevE.71.030102} {\bibfield  {journal} {\bibinfo  {journal}
  {Physical Review E}\ }\textbf {\bibinfo {volume} {71}},\ \bibinfo {pages}
  {030102(R)} (\bibinfo {year} {2005})}\BibitemShut {NoStop}%
\bibitem [{\citenamefont {Sasa}(2012)}]{Sasa2012}%
  \BibitemOpen
  \bibfield  {author} {\bibinfo {author} {\bibfnamefont {S.-i.}\ \bibnamefont
  {Sasa}},\ }\href {\doibase 10.1103/PhysRevLett.109.165702} {\bibfield
  {journal} {\bibinfo  {journal} {Physical Review Letters}\ }\textbf {\bibinfo
  {volume} {109}},\ \bibinfo {pages} {165702} (\bibinfo {year}
  {2012})}\BibitemShut {NoStop}%
\bibitem [{\citenamefont {Wang}\ and\ \citenamefont
  {Landau}(2001{\natexlab{a}})}]{Wang2001}%
  \BibitemOpen
  \bibfield  {author} {\bibinfo {author} {\bibfnamefont {F.}~\bibnamefont
  {Wang}}\ and\ \bibinfo {author} {\bibfnamefont {D.~P.}\ \bibnamefont
  {Landau}},\ }\href {\doibase 10.1103/PhysRevLett.86.2050} {\bibfield
  {journal} {\bibinfo  {journal} {Physical Review Letters}\ }\textbf {\bibinfo
  {volume} {86}},\ \bibinfo {pages} {2050} (\bibinfo {year}
  {2001}{\natexlab{a}})}\BibitemShut {NoStop}%
\bibitem [{\citenamefont {Wang}\ and\ \citenamefont
  {Landau}(2001{\natexlab{b}})}]{Wang2001a}%
  \BibitemOpen
  \bibfield  {author} {\bibinfo {author} {\bibfnamefont {F.}~\bibnamefont
  {Wang}}\ and\ \bibinfo {author} {\bibfnamefont {D.~P.}\ \bibnamefont
  {Landau}},\ }\href {\doibase 10.1103/PhysRevE.64.056101} {\bibfield
  {journal} {\bibinfo  {journal} {Physical Review E}\ }\textbf {\bibinfo
  {volume} {64}},\ \bibinfo {pages} {056101} (\bibinfo {year}
  {2001}{\natexlab{b}})}\BibitemShut {NoStop}%
\bibitem [{\citenamefont {Franz}\ and\ \citenamefont
  {Parisi}(1995)}]{Franz1995}%
  \BibitemOpen
  \bibfield  {author} {\bibinfo {author} {\bibfnamefont {S.}~\bibnamefont
  {Franz}}\ and\ \bibinfo {author} {\bibfnamefont {G.}~\bibnamefont {Parisi}},\
  }\href {\doibase 10.1051/jp1:1995201} {\bibfield  {journal} {\bibinfo
  {journal} {Journal de Physique I}\ ,\ \bibinfo {pages} {21}} (\bibinfo {year}
  {1995})}\BibitemShut {NoStop}%
\bibitem [{\citenamefont {Franz}\ and\ \citenamefont
  {Parisi}(1997)}]{Franz1997}%
  \BibitemOpen
  \bibfield  {author} {\bibinfo {author} {\bibfnamefont {S.}~\bibnamefont
  {Franz}}\ and\ \bibinfo {author} {\bibfnamefont {G.}~\bibnamefont {Parisi}},\
  }\href {\doibase 10.1103/PhysRevLett.79.2486} {\bibfield  {journal} {\bibinfo
   {journal} {Physical Review Letters}\ }\textbf {\bibinfo {volume} {79}},\
  \bibinfo {pages} {2486} (\bibinfo {year} {1997})}\BibitemShut {NoStop}%
\bibitem [{\citenamefont {Franz}\ and\ \citenamefont
  {Parisi}(1998)}]{Franz1998}%
  \BibitemOpen
  \bibfield  {author} {\bibinfo {author} {\bibfnamefont {S.}~\bibnamefont
  {Franz}}\ and\ \bibinfo {author} {\bibfnamefont {G.}~\bibnamefont {Parisi}},\
  }\href {http://www.sciencedirect.com/science/article/pii/S037843719800315X}
  {\bibfield  {journal} {\bibinfo  {journal} {Physica A: Statistical Mechanics
  and its Applications}\ }\textbf {\bibinfo {volume} {261}},\ \bibinfo {pages}
  {317} (\bibinfo {year} {1998})}\BibitemShut {NoStop}%
\bibitem [{\citenamefont {Kawasaki}(1966)}]{Kawasaki1966}%
  \BibitemOpen
  \bibfield  {author} {\bibinfo {author} {\bibfnamefont {K.}~\bibnamefont
  {Kawasaki}},\ }\href@noop {} {\bibfield  {journal} {\bibinfo  {journal}
  {Phys. Rev.}\ }\textbf {\bibinfo {volume} {145}},\ \bibinfo {pages} {224}
  (\bibinfo {year} {1966})}\BibitemShut {NoStop}%
\bibitem [{\citenamefont {Foini}\ \emph {et~al.}(2011)\citenamefont {Foini},
  \citenamefont {Semerjian},\ and\ \citenamefont {Zamponi}}]{Foini2011}%
  \BibitemOpen
  \bibfield  {author} {\bibinfo {author} {\bibfnamefont {L.}~\bibnamefont
  {Foini}}, \bibinfo {author} {\bibfnamefont {G.}~\bibnamefont {Semerjian}}, \
  and\ \bibinfo {author} {\bibfnamefont {F.}~\bibnamefont {Zamponi}},\ }\href
  {\doibase 10.1103/PhysRevB.83.094513} {\bibfield  {journal} {\bibinfo
  {journal} {Physical Review B}\ }\textbf {\bibinfo {volume} {83}},\ \bibinfo
  {pages} {094513} (\bibinfo {year} {2011})}\BibitemShut {NoStop}%
\bibitem [{Sup()}]{Supplement}%
  \BibitemOpen
  \href@noop {} {\enquote {\bibinfo {title} {See {Supplemental Material} for a
  further discussion on the dynamics, the finite-size dependence of the
  relaxation time, and the determination of the dynamical transition
  temperature {$T_\mathrm d$}, which includes
  {Refs.}~\cite{Berthier2011,Berthier2012,Biroli2001,Coslovich2019,Franz1997,Franz1998,Franz2013,Karmakar2009}},}\
  }\BibitemShut {NoStop}%
\bibitem [{\citenamefont {Kob}\ \emph {et~al.}(1997)\citenamefont {Kob},
  \citenamefont {Donati}, \citenamefont {Plimpton}, \citenamefont {Poole},\
  and\ \citenamefont {Glotzer}}]{Kob1997}%
  \BibitemOpen
  \bibfield  {author} {\bibinfo {author} {\bibfnamefont {W.}~\bibnamefont
  {Kob}}, \bibinfo {author} {\bibfnamefont {C.}~\bibnamefont {Donati}},
  \bibinfo {author} {\bibfnamefont {S.~J.}\ \bibnamefont {Plimpton}}, \bibinfo
  {author} {\bibfnamefont {P.~H.}\ \bibnamefont {Poole}}, \ and\ \bibinfo
  {author} {\bibfnamefont {S.~C.}\ \bibnamefont {Glotzer}},\ }\href {\doibase
  10.1103/PhysRevLett.79.2827} {\bibfield  {journal} {\bibinfo  {journal}
  {Physical Review Letters}\ }\textbf {\bibinfo {volume} {79}},\ \bibinfo
  {pages} {2827} (\bibinfo {year} {1997})}\BibitemShut {NoStop}%
\bibitem [{\citenamefont {Yamamoto}\ and\ \citenamefont
  {Onuki}(1998)}]{Yamamoto1998}%
  \BibitemOpen
  \bibfield  {author} {\bibinfo {author} {\bibfnamefont {R.}~\bibnamefont
  {Yamamoto}}\ and\ \bibinfo {author} {\bibfnamefont {A.}~\bibnamefont
  {Onuki}},\ }\href {\doibase 10.1103/PhysRevLett.81.4915} {\bibfield
  {journal} {\bibinfo  {journal} {Physical Review Letters}\ }\textbf {\bibinfo
  {volume} {81}},\ \bibinfo {pages} {4915} (\bibinfo {year}
  {1998})}\BibitemShut {NoStop}%
\bibitem [{\citenamefont {Ediger}(2000)}]{Ediger2000}%
  \BibitemOpen
  \bibfield  {author} {\bibinfo {author} {\bibfnamefont {M.~D.}\ \bibnamefont
  {Ediger}},\ }\href {\doibase 10.1146/annurev.physchem.51.1.99} {\bibfield
  {journal} {\bibinfo  {journal} {Annual Review of Physical Chemistry}\
  }\textbf {\bibinfo {volume} {51}},\ \bibinfo {pages} {99} (\bibinfo {year}
  {2000})}\BibitemShut {NoStop}%
\bibitem [{\citenamefont {Toninelli}\ \emph {et~al.}(2005)\citenamefont
  {Toninelli}, \citenamefont {Wyart}, \citenamefont {Berthier}, \citenamefont
  {Biroli},\ and\ \citenamefont {Bouchaud}}]{Toninelli2005}%
  \BibitemOpen
  \bibfield  {author} {\bibinfo {author} {\bibfnamefont {C.}~\bibnamefont
  {Toninelli}}, \bibinfo {author} {\bibfnamefont {M.}~\bibnamefont {Wyart}},
  \bibinfo {author} {\bibfnamefont {L.}~\bibnamefont {Berthier}}, \bibinfo
  {author} {\bibfnamefont {G.}~\bibnamefont {Biroli}}, \ and\ \bibinfo {author}
  {\bibfnamefont {J.-P.}\ \bibnamefont {Bouchaud}},\ }\href {\doibase
  10.1103/PhysRevE.71.041505} {\bibfield  {journal} {\bibinfo  {journal}
  {Physical Review E}\ }\textbf {\bibinfo {volume} {71}},\ \bibinfo {pages}
  {041505} (\bibinfo {year} {2005})}\BibitemShut {NoStop}%
\bibitem [{\citenamefont {Cammarota}\ and\ \citenamefont
  {Biroli}(2012)}]{Cammarota2012}%
  \BibitemOpen
  \bibfield  {author} {\bibinfo {author} {\bibfnamefont {C.}~\bibnamefont
  {Cammarota}}\ and\ \bibinfo {author} {\bibfnamefont {G.}~\bibnamefont
  {Biroli}},\ }\href {\doibase 10.1073/pnas.1111582109} {\bibfield  {journal}
  {\bibinfo  {journal} {Proceedings of the National Academy of Sciences}\
  }\textbf {\bibinfo {volume} {109}},\ \bibinfo {pages} {8850} (\bibinfo {year}
  {2012})}\BibitemShut {NoStop}%
\bibitem [{\citenamefont {Cammarota}\ and\ \citenamefont
  {Biroli}(2013)}]{Cammarota2013}%
  \BibitemOpen
  \bibfield  {author} {\bibinfo {author} {\bibfnamefont {C.}~\bibnamefont
  {Cammarota}}\ and\ \bibinfo {author} {\bibfnamefont {G.}~\bibnamefont
  {Biroli}},\ }\href {\doibase 10.1063/1.4790400} {\bibfield  {journal}
  {\bibinfo  {journal} {The Journal of chemical physics}\ }\textbf {\bibinfo
  {volume} {138}},\ \bibinfo {pages} {12A547} (\bibinfo {year}
  {2013})}\BibitemShut {NoStop}%
\bibitem [{\citenamefont {Kob}\ and\ \citenamefont {Berthier}(2013)}]{Kob2013}%
  \BibitemOpen
  \bibfield  {author} {\bibinfo {author} {\bibfnamefont {W.}~\bibnamefont
  {Kob}}\ and\ \bibinfo {author} {\bibfnamefont {L.}~\bibnamefont {Berthier}},\
  }\href {\doibase 10.1103/PhysRevLett.110.245702} {\bibfield  {journal}
  {\bibinfo  {journal} {Physical Review Letters}\ }\textbf {\bibinfo {volume}
  {110}},\ \bibinfo {pages} {245702} (\bibinfo {year} {2013})}\BibitemShut
  {NoStop}%
\bibitem [{\citenamefont {Ozawa}\ \emph {et~al.}(2015)\citenamefont {Ozawa},
  \citenamefont {Kob}, \citenamefont {Ikeda},\ and\ \citenamefont
  {Miyazaki}}]{Ozawa2015}%
  \BibitemOpen
  \bibfield  {author} {\bibinfo {author} {\bibfnamefont {M.}~\bibnamefont
  {Ozawa}}, \bibinfo {author} {\bibfnamefont {W.}~\bibnamefont {Kob}}, \bibinfo
  {author} {\bibfnamefont {A.}~\bibnamefont {Ikeda}}, \ and\ \bibinfo {author}
  {\bibfnamefont {K.}~\bibnamefont {Miyazaki}},\ }\href {\doibase
  10.1073/pnas.1500730112} {\bibfield  {journal} {\bibinfo  {journal}
  {Proceedings of the National Academy of Sciences}\ }\textbf {\bibinfo
  {volume} {112}},\ \bibinfo {pages} {6914} (\bibinfo {year}
  {2015})}\BibitemShut {NoStop}%
\bibitem [{\citenamefont {M{\'e}zard}\ and\ \citenamefont
  {Parisi}(2012)}]{Mezard2012}%
  \BibitemOpen
  \bibfield  {author} {\bibinfo {author} {\bibfnamefont {M.}~\bibnamefont
  {M{\'e}zard}}\ and\ \bibinfo {author} {\bibfnamefont {G.}~\bibnamefont
  {Parisi}},\ }\enquote {\bibinfo {title} {Glasses and replicas},}\ in\ \href
  {\doibase 10.1002/9781118202470.ch4} {\emph {\bibinfo {booktitle} {Structural
  Glasses and Supercooled Liquids}}}\ (\bibinfo  {publisher} {John Wiley \&
  Sons, Ltd},\ \bibinfo {year} {2012})\ Chap.~\bibinfo {chapter} {4}, pp.\
  \bibinfo {pages} {151--191}\BibitemShut {NoStop}%
\bibitem [{\citenamefont {Takahashi}\ and\ \citenamefont
  {Hukushima}(2015)}]{Takahashi2015}%
  \BibitemOpen
  \bibfield  {author} {\bibinfo {author} {\bibfnamefont {T.}~\bibnamefont
  {Takahashi}}\ and\ \bibinfo {author} {\bibfnamefont {K.}~\bibnamefont
  {Hukushima}},\ }\href {\doibase 10.1103/PhysRevE.91.020102} {\bibfield
  {journal} {\bibinfo  {journal} {Physical Review E}\ }\textbf {\bibinfo
  {volume} {91}},\ \bibinfo {pages} {020102(R)} (\bibinfo {year}
  {2015})}\BibitemShut {NoStop}%
\bibitem [{\citenamefont {Takahashi}\ and\ \citenamefont
  {Hukushima}(shed)}]{TakahashiHukushima}%
  \BibitemOpen
  \bibfield  {author} {\bibinfo {author} {\bibfnamefont {T.}~\bibnamefont
  {Takahashi}}\ and\ \bibinfo {author} {\bibfnamefont {K.}~\bibnamefont
  {Hukushima}},\ }\href@noop {} {} (\bibinfo {year} {unpublished})\BibitemShut
  {NoStop}%
\bibitem [{\citenamefont {Franz}\ and\ \citenamefont
  {Parisi}(2013)}]{Franz2013}%
  \BibitemOpen
  \bibfield  {author} {\bibinfo {author} {\bibfnamefont {S.}~\bibnamefont
  {Franz}}\ and\ \bibinfo {author} {\bibfnamefont {G.}~\bibnamefont {Parisi}},\
  }\href {\doibase 10.1088/1742-5468/2013/11/P11012} {\bibfield  {journal}
  {\bibinfo  {journal} {Journal of Statistical Mechanics: Theory and
  Experiment}\ }\textbf {\bibinfo {volume} {2013}},\ \bibinfo {pages} {P11012}
  (\bibinfo {year} {2013})}\BibitemShut {NoStop}%
\bibitem [{\citenamefont {Berthier}\ and\ \citenamefont
  {Coslovich}(2014)}]{Berthier2014}%
  \BibitemOpen
  \bibfield  {author} {\bibinfo {author} {\bibfnamefont {L.}~\bibnamefont
  {Berthier}}\ and\ \bibinfo {author} {\bibfnamefont {D.}~\bibnamefont
  {Coslovich}},\ }\href {\doibase 10.1073/pnas.1407934111} {\bibfield
  {journal} {\bibinfo  {journal} {Proceedings of the National Academy of
  Sciences}\ }\textbf {\bibinfo {volume} {111}},\ \bibinfo {pages} {11668}
  (\bibinfo {year} {2014})}\BibitemShut {NoStop}%
\bibitem [{\citenamefont {Berthier}\ \emph {et~al.}(2017)\citenamefont
  {Berthier}, \citenamefont {Charbonneau}, \citenamefont {Coslovich},
  \citenamefont {Ninarello}, \citenamefont {Ozawa},\ and\ \citenamefont
  {Yaida}}]{Berthier2017}%
  \BibitemOpen
  \bibfield  {author} {\bibinfo {author} {\bibfnamefont {L.}~\bibnamefont
  {Berthier}}, \bibinfo {author} {\bibfnamefont {P.}~\bibnamefont
  {Charbonneau}}, \bibinfo {author} {\bibfnamefont {D.}~\bibnamefont
  {Coslovich}}, \bibinfo {author} {\bibfnamefont {A.}~\bibnamefont
  {Ninarello}}, \bibinfo {author} {\bibfnamefont {M.}~\bibnamefont {Ozawa}}, \
  and\ \bibinfo {author} {\bibfnamefont {S.}~\bibnamefont {Yaida}},\ }\href
  {\doibase 10.1073/pnas.1706860114} {\bibfield  {journal} {\bibinfo  {journal}
  {Proceedings of the National Academy of Sciences}\ }\textbf {\bibinfo
  {volume} {114}},\ \bibinfo {pages} {11356} (\bibinfo {year}
  {2017})}\BibitemShut {NoStop}%
\bibitem [{\citenamefont {Berg}\ and\ \citenamefont {Janke}(1998)}]{Berg1998}%
  \BibitemOpen
  \bibfield  {author} {\bibinfo {author} {\bibfnamefont {B.~A.}\ \bibnamefont
  {Berg}}\ and\ \bibinfo {author} {\bibfnamefont {W.}~\bibnamefont {Janke}},\
  }\href {\doibase 10.1103/PhysRevLett.80.4771} {\bibfield  {journal} {\bibinfo
   {journal} {Physical Review Letters}\ }\textbf {\bibinfo {volume} {80}},\
  \bibinfo {pages} {4771} (\bibinfo {year} {1998})}\BibitemShut {NoStop}%
\bibitem [{Note1()}]{Note1}%
  \BibitemOpen
  \bibinfo {note} {Note that the Franz--Parisi potential is convex at any
  temperature in the thermodynamic limit as the free-energy barrier is
  sub-extensive, and the non-convexity is seen only in finite
  systems.}\BibitemShut {Stop}%
\bibitem [{\citenamefont {Berthier}(2013)}]{Berthier2013}%
  \BibitemOpen
  \bibfield  {author} {\bibinfo {author} {\bibfnamefont {L.}~\bibnamefont
  {Berthier}},\ }\href {\doibase 10.1103/PhysRevE.88.022313} {\bibfield
  {journal} {\bibinfo  {journal} {Physical Review E}\ }\textbf {\bibinfo
  {volume} {88}},\ \bibinfo {pages} {022313} (\bibinfo {year}
  {2013})}\BibitemShut {NoStop}%
\bibitem [{\citenamefont {Berthier}\ and\ \citenamefont
  {Jack}(2015)}]{Berthier2015}%
  \BibitemOpen
  \bibfield  {author} {\bibinfo {author} {\bibfnamefont {L.}~\bibnamefont
  {Berthier}}\ and\ \bibinfo {author} {\bibfnamefont {R.~L.}\ \bibnamefont
  {Jack}},\ }\href {\doibase 10.1103/PhysRevLett.114.205701} {\bibfield
  {journal} {\bibinfo  {journal} {Physical Review Letters}\ }\textbf {\bibinfo
  {volume} {114}},\ \bibinfo {pages} {205701} (\bibinfo {year}
  {2015})}\BibitemShut {NoStop}%
\bibitem [{\citenamefont {Fisher}\ and\ \citenamefont
  {Berker}(1982)}]{Fisher1982}%
  \BibitemOpen
  \bibfield  {author} {\bibinfo {author} {\bibfnamefont {M.~E.}\ \bibnamefont
  {Fisher}}\ and\ \bibinfo {author} {\bibfnamefont {A.~N.}\ \bibnamefont
  {Berker}},\ }\href {\doibase 10.1103/PhysRevB.26.2507} {\bibfield  {journal}
  {\bibinfo  {journal} {Physical Review B}\ }\textbf {\bibinfo {volume} {26}},\
  \bibinfo {pages} {2507} (\bibinfo {year} {1982})}\BibitemShut {NoStop}%
\bibitem [{\citenamefont {Mueller}\ \emph {et~al.}(2014)\citenamefont
  {Mueller}, \citenamefont {Janke},\ and\ \citenamefont
  {Johnston}}]{Mueller2014}%
  \BibitemOpen
  \bibfield  {author} {\bibinfo {author} {\bibfnamefont {M.}~\bibnamefont
  {Mueller}}, \bibinfo {author} {\bibfnamefont {W.}~\bibnamefont {Janke}}, \
  and\ \bibinfo {author} {\bibfnamefont {D.~A.}\ \bibnamefont {Johnston}},\
  }\href {\doibase 10.1103/PhysRevLett.112.200601} {\bibfield  {journal}
  {\bibinfo  {journal} {Physical Review Letters}\ }\textbf {\bibinfo {volume}
  {112}},\ \bibinfo {pages} {200601} (\bibinfo {year} {2014})}\BibitemShut
  {NoStop}%
\bibitem [{\citenamefont {Hukushima}\ and\ \citenamefont
  {Nemoto}(1996)}]{Hukushima1996}%
  \BibitemOpen
  \bibfield  {author} {\bibinfo {author} {\bibfnamefont {K.}~\bibnamefont
  {Hukushima}}\ and\ \bibinfo {author} {\bibfnamefont {K.}~\bibnamefont
  {Nemoto}},\ }\href {\doibase 10.1143/JPSJ.65.1604} {\bibfield  {journal}
  {\bibinfo  {journal} {Journal of the Physical Society of Japan}\ }\textbf
  {\bibinfo {volume} {65}},\ \bibinfo {pages} {1604} (\bibinfo {year}
  {1996})}\BibitemShut {NoStop}%
\bibitem [{\citenamefont {Ferrenberg}\ and\ \citenamefont
  {Swendsen}(1988)}]{Ferrenberg1988}%
  \BibitemOpen
  \bibfield  {author} {\bibinfo {author} {\bibfnamefont {A.~M.}\ \bibnamefont
  {Ferrenberg}}\ and\ \bibinfo {author} {\bibfnamefont {R.~H.}\ \bibnamefont
  {Swendsen}},\ }\href {\doibase 10.1103/PhysRevLett.61.2635} {\bibfield
  {journal} {\bibinfo  {journal} {Physical Review Letters}\ }\textbf {\bibinfo
  {volume} {61}},\ \bibinfo {pages} {2635} (\bibinfo {year}
  {1988})}\BibitemShut {NoStop}%
\bibitem [{\citenamefont {Ferrenberg}\ and\ \citenamefont
  {Swendsen}(1989)}]{Ferrenberg1989}%
  \BibitemOpen
  \bibfield  {author} {\bibinfo {author} {\bibfnamefont {A.~M.}\ \bibnamefont
  {Ferrenberg}}\ and\ \bibinfo {author} {\bibfnamefont {R.~H.}\ \bibnamefont
  {Swendsen}},\ }\href {\doibase 10.1103/PhysRevLett.63.1195} {\bibfield
  {journal} {\bibinfo  {journal} {Physical Review Letters}\ }\textbf {\bibinfo
  {volume} {63}},\ \bibinfo {pages} {1195} (\bibinfo {year}
  {1989})}\BibitemShut {NoStop}%
\bibitem [{\citenamefont {M{\"{u}}nger}\ and\ \citenamefont
  {Novotny}(1991)}]{Munger1991}%
  \BibitemOpen
  \bibfield  {author} {\bibinfo {author} {\bibfnamefont {E.~P.}\ \bibnamefont
  {M{\"{u}}nger}}\ and\ \bibinfo {author} {\bibfnamefont {M.~A.}\ \bibnamefont
  {Novotny}},\ }\href {\doibase 10.1103/PhysRevB.43.5773} {\bibfield  {journal}
  {\bibinfo  {journal} {Physical Review B}\ }\textbf {\bibinfo {volume} {43}},\
  \bibinfo {pages} {5773} (\bibinfo {year} {1991})}\BibitemShut {NoStop}%
\bibitem [{\citenamefont {Billoire}\ \emph {et~al.}(2005)\citenamefont
  {Billoire}, \citenamefont {Giomi},\ and\ \citenamefont
  {Marinari}}]{Billoire2005}%
  \BibitemOpen
  \bibfield  {author} {\bibinfo {author} {\bibfnamefont {A.}~\bibnamefont
  {Billoire}}, \bibinfo {author} {\bibfnamefont {L.}~\bibnamefont {Giomi}}, \
  and\ \bibinfo {author} {\bibfnamefont {E.}~\bibnamefont {Marinari}},\ }\href
  {\doibase 10.1209/epl/i2005-10149-4} {\bibfield  {journal} {\bibinfo
  {journal} {Europhysics Letters (EPL)}\ }\textbf {\bibinfo {volume} {71}},\
  \bibinfo {pages} {824} (\bibinfo {year} {2005})}\BibitemShut {NoStop}%
\bibitem [{\citenamefont {Berthier}\ \emph {et~al.}(2012)\citenamefont
  {Berthier}, \citenamefont {Biroli}, \citenamefont {Coslovich}, \citenamefont
  {Kob},\ and\ \citenamefont {Toninelli}}]{Berthier2012}%
  \BibitemOpen
  \bibfield  {author} {\bibinfo {author} {\bibfnamefont {L.}~\bibnamefont
  {Berthier}}, \bibinfo {author} {\bibfnamefont {G.}~\bibnamefont {Biroli}},
  \bibinfo {author} {\bibfnamefont {D.}~\bibnamefont {Coslovich}}, \bibinfo
  {author} {\bibfnamefont {W.}~\bibnamefont {Kob}}, \ and\ \bibinfo {author}
  {\bibfnamefont {C.}~\bibnamefont {Toninelli}},\ }\href {\doibase
  10.1103/PhysRevE.86.031502} {\bibfield  {journal} {\bibinfo  {journal}
  {Physical Review E}\ }\textbf {\bibinfo {volume} {86}},\ \bibinfo {pages}
  {031502} (\bibinfo {year} {2012})}\BibitemShut {NoStop}%
\bibitem [{\citenamefont {Coslovich}\ \emph {et~al.}(2019)\citenamefont
  {Coslovich}, \citenamefont {Ninarello},\ and\ \citenamefont
  {Berthier}}]{Coslovich2019}%
  \BibitemOpen
  \bibfield  {author} {\bibinfo {author} {\bibfnamefont {D.}~\bibnamefont
  {Coslovich}}, \bibinfo {author} {\bibfnamefont {A.}~\bibnamefont
  {Ninarello}}, \ and\ \bibinfo {author} {\bibfnamefont {L.}~\bibnamefont
  {Berthier}},\ }\href {\doibase 10.21468/SciPostPhys.7.6.077} {\bibfield
  {journal} {\bibinfo  {journal} {SciPost Physics}\ }\textbf {\bibinfo {volume}
  {7}},\ \bibinfo {pages} {077} (\bibinfo {year} {2019})},\ \Eprint
  {http://arxiv.org/abs/1811.03171} {1811.03171} \BibitemShut {NoStop}%
\bibitem [{\citenamefont {Karmakar}\ \emph {et~al.}(2009)\citenamefont
  {Karmakar}, \citenamefont {Dasgupta},\ and\ \citenamefont
  {Sastry}}]{Karmakar2009}%
  \BibitemOpen
  \bibfield  {author} {\bibinfo {author} {\bibfnamefont {S.}~\bibnamefont
  {Karmakar}}, \bibinfo {author} {\bibfnamefont {C.}~\bibnamefont {Dasgupta}},
  \ and\ \bibinfo {author} {\bibfnamefont {S.}~\bibnamefont {Sastry}},\ }\href
  {\doibase 10.1073/pnas.0811082106} {\bibfield  {journal} {\bibinfo  {journal}
  {Proceedings of the National Academy of Sciences of the United States of
  America}\ }\textbf {\bibinfo {volume} {106}},\ \bibinfo {pages} {3675}
  (\bibinfo {year} {2009})}\BibitemShut {NoStop}%
\end{thebibliography}%
\nocite{Berthier2011,Berthier2012,Biroli2001,Coslovich2019,Franz1997,Franz1998,Franz2013,Karmakar2009}

\ifincludesupplements

\clearpage
\title{Supplemental Material for ``Lattice glass model in three spatial dimensions''}
\author{Yoshihiko Nishikawa}
\author{Koji Hukushima}

\maketitle

\onecolumngrid
\renewcommand{\theequation}{S\arabic{equation}}
\renewcommand{\thefigure}{S\arabic{figure}}
\setcounter{equation}{0}
\setcounter{figure}{0}
\setcounter{page}{1}
\setcounter{section}{0}

\subsection{Supplemental Item 1: Dynamics at low temperature}
\label{sec:dynamics_plateau}
In off-lattice particle models for supercooled liquids and glasses, the dynamics at low
temperature shows two-step relaxation as well as our model. The physical interpretation 
of the plateau regime in the particle models is well known that particles vibrate locally 
inside cages effectively formed by surrounding particles \cite{Berthier2011}. However, since 
particles of our model 
defined on a lattice have discrete degrees of freedom, the appropriate physical
interpretation analogous to the local vibration 
in the plateau regime is unclear. In this supplemental item,
we show evidence of local vibrations of particles during the plateau regime.

Thanks to the discrete nature of our model, it is easy to find lattice sites that have 
no contribution to the autocorrelation function Eq.~(2), 
where $\delta_{\sigma_i \pt{t_\mathrm w}, \sigma_i \pt{t_\mathrm w + t}} = 0$, i.e.
the lattice sites that have different types of particles at $t = 0$ and $t > 0$ or 
that are vacant at $t > 0$ but occupied at $t = 0$.
Roughly speaking, a cluster represents a movable region during time
interval $t$. 
We identify connected clusters 
of those lattice sites at time $t$, and compute the normalized distribution of the 
cluster size $n_c$.
\Fig{fig:cluster_hist} shows the distribution of the connected-cluster
size depending on the time $t$ . 

At temperature $T / A = 0.35$, the autocorrelation function has 
a plateau from time $t \simeq 10$ to $t \simeq 10^3$, see Fig~2.
In this time regime, the size of connected clusters $n_\mathrm c \lesssim 10$, 
and the distribution is almost independent of time, see \Fig{fig:cluster_hist}.
Although the size distribution does not depend on time very much, the positions 
of the clusters change with time during the plateau, and some of them disappear 
with time (see \Fig{fig:visualization} for real-space visualization 
of the changed lattice sites). We thus conclude, at short time scale $t \lesssim 10^3$, 
the system has local vibrations that involve only $O(1)$ number of particles.
These local vibrations produce the plateau in the autocorrelation
function, even in the lattice model. 

As time passes, the distribution of $n_\mathrm c$ has a broad tail towards 
larger $n_\mathrm c$, and eventually becomes double-peaked at large time 
$t \simeq \tau_\alpha \simeq 10^7$. The emergence of two peaks in the distribution
with small clusters of $n_\mathrm c = O(1)$ and giant clusters of $n_\mathrm c = O(N)$ 
indicates the heterogeneous dynamics. At time $t = 10^7$, the largest 
cluster spreads over the system while there are still some regions where only small 
clusters occupy, see \Fig{fig:visualization}.

\begin{figure}[htbp]
\center \includegraphics[width=.4\linewidth]{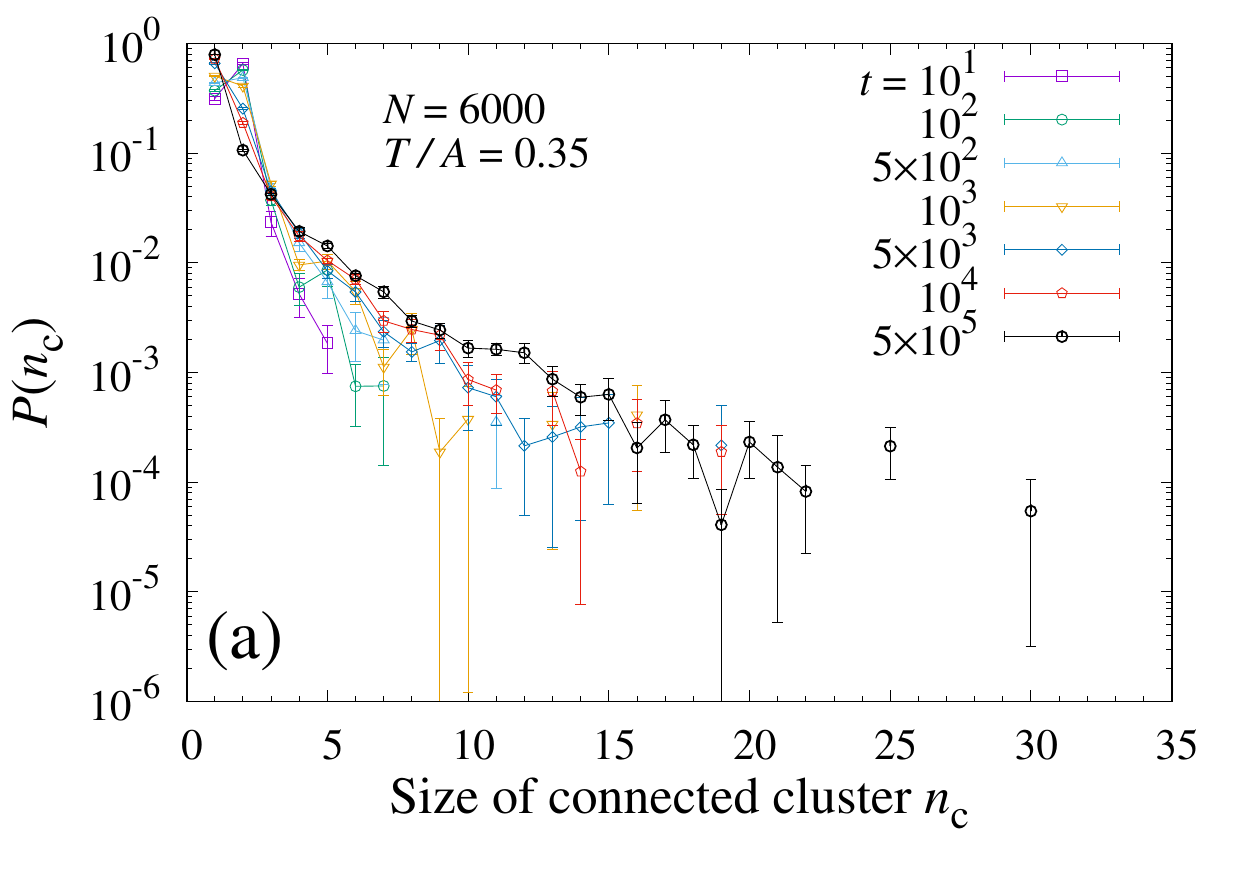}%
\includegraphics[width=.4\linewidth]{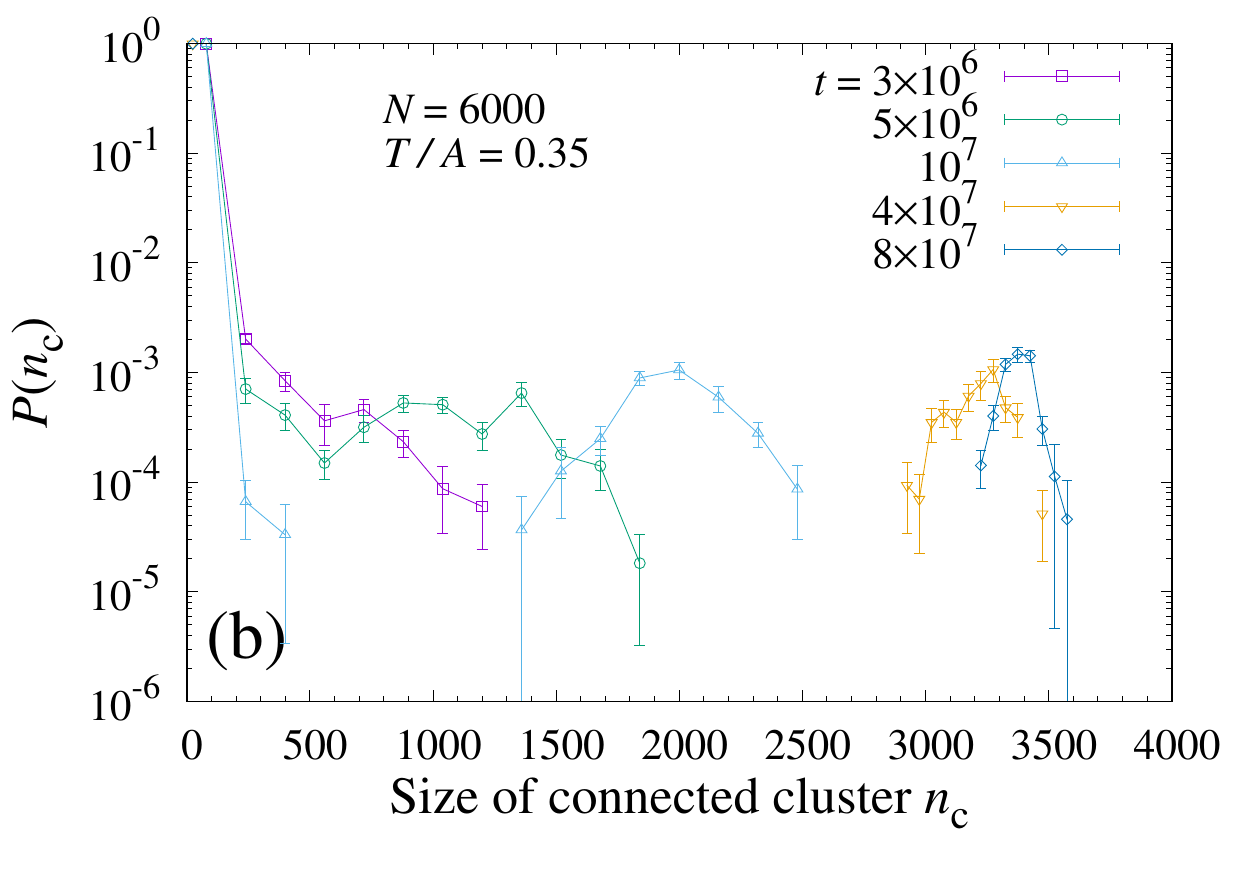}%
 \caption{
The normalized distribution function of the size of connected clusters at $T / A = 0.35$
for (a) $10^1 \leq t \leq 5 \times 10^5$ and (b) $3 \times 10^6 \leq t \leq 8 \times 10^7$.
The system size $L = 20$ ($N = 6000$). 
}
\label{fig:cluster_hist}
\end{figure}

\begin{figure}[htbp]
\includegraphics[width=.3\linewidth]{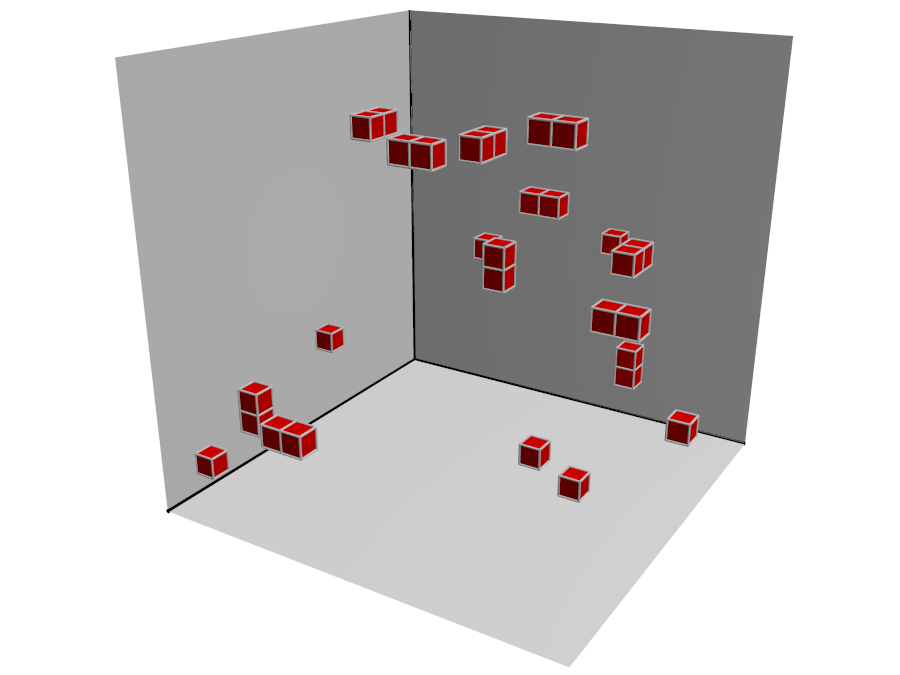}%
\includegraphics[width=.3\linewidth]{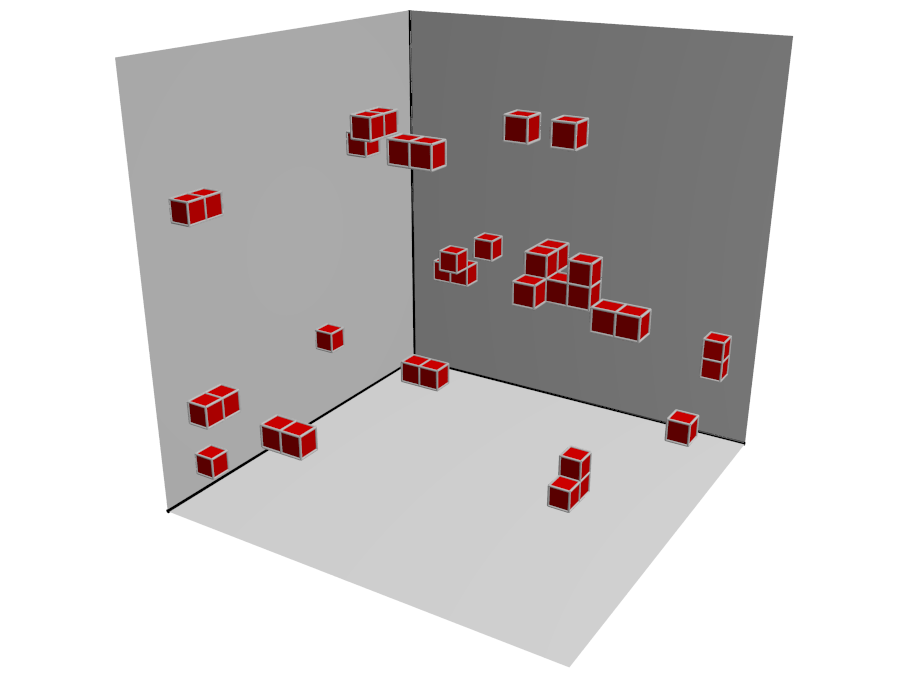}%
\includegraphics[width=.3\linewidth]{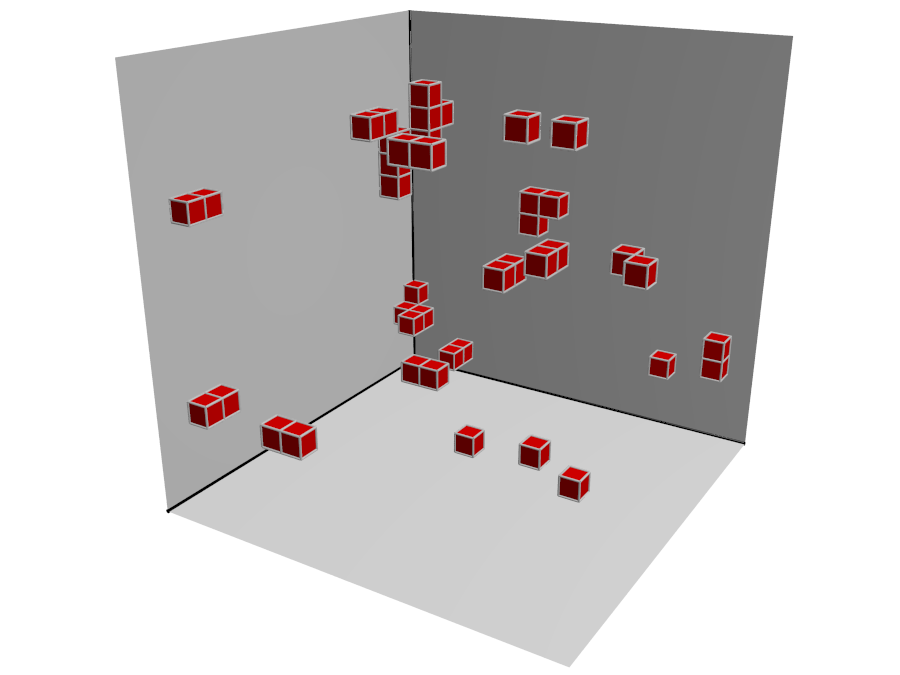}
\includegraphics[width=.3\linewidth]{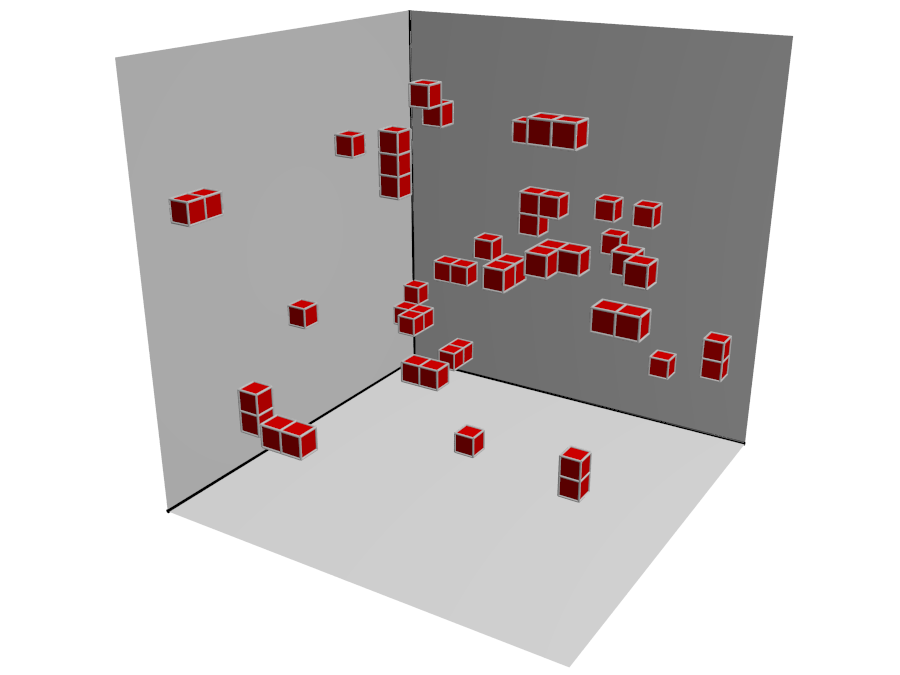}%
\includegraphics[width=.3\linewidth]{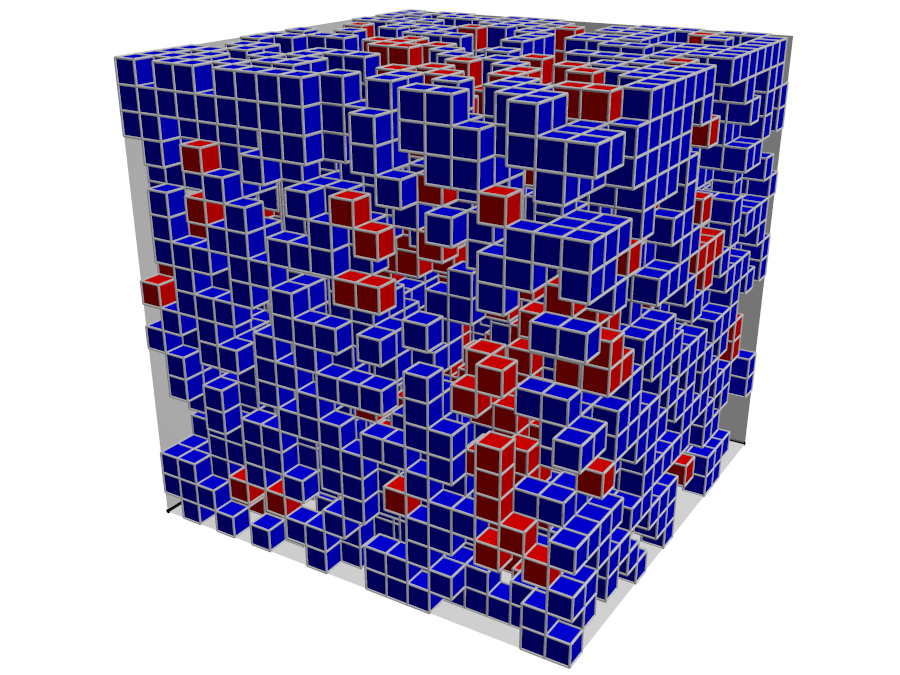}%
\caption{
Real-space visualization of typical configurations at time $t = 10^1, 10^2, 5\times10^2, 10^3$,
and $10^7$, from left top to right bottom. The temperature $T / A = 0.35$. The lattice sites 
which have different particles in the configuration at finite time $t > 0$ from the original 
configuration at $t = 0$ are filled with cubes. Here, at this temperature, the autocorrelation 
function has a plateau when $10^1 \lesssim t \lesssim 10^3$. For the configuration at $t = 10^7$, 
the lattice sites belonging to the largest connected cluster are filled with blue cubes, and those 
for the other clusters are filled with red cubes. 
}
\label{fig:visualization}
\end{figure}

\subsection{Supplemental Item 2: Estimation of the dynamical transition temperature}

In the RFOT theory, the relaxation time diverges algebraically at temperature 
$T_\mathrm d >T_\mathrm K$ rather than exponentially as observed in experiments. 
Whereas the dynamical transition turns into a crossover in finite dimensions, 
a well-defined localization transition of the potential energy landscape
was shown to control the dynamical cross-over \cite{Coslovich2019}.
However, their method to identify the localization transition is not available in our 
model. We thus estimate the dynamical transition temperature $T_\mathrm d$ by a 
more traditional method, that is, power-law fitting of the relaxation time 
as done in Ref.~\cite{Biroli2001}.

We show in \Fig{fig:Tc} the best fit of the inverse relaxation time to a power-law behavior 
$\sim (T - T_\mathrm d)^{\gamma}$, with $T_\mathrm d = 0.348(4)$ and $\gamma = 4.18(12)$.
In mean-field models, the dynamical transition temperature is close to the 
critical temperature of the $\varepsilon$-coupled system \cite{Franz1997,Franz1998,Franz2013}.
Our estimation of $T_\mathrm d$ is compatible with the temperature-coupling phase diagram
Fig.~6. 

\begin{figure}[htbp]
\center \includegraphics[width=.5\linewidth]{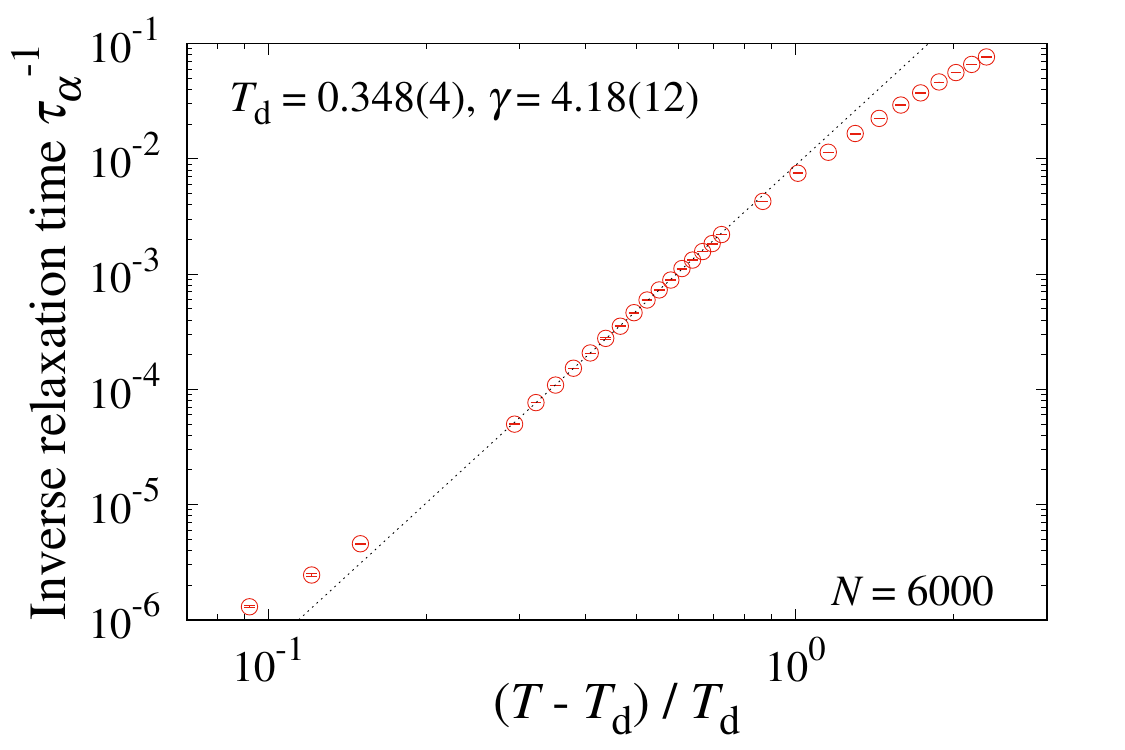}%
 \caption{
The inverse relaxation time $\tau_\alpha^{-1}$ as a function of $(T - T_\mathrm d) 
/ T_\mathrm d$ and a power-law curve $\sim (T - T_\mathrm d)^{\gamma}$.
The dynamical transition temperature $T_\mathrm c$ and the exponent $\gamma$ are 
estimated by fitting using the data at $0.4 \leq T / A \leq 0.6$, 
as $T_\mathrm d = 0.348(4)$ and $\gamma = 4.18(12)$, respectively. 
}
\label{fig:Tc}
\end{figure}

\subsection{Supplemental Item 3: Finite-size effect in the relaxation time}

The relaxation time of glass forming liquids has a decreasing behavior 
at moderately low temperature \cite{Karmakar2009,Berthier2012}. 
The magnitude of the decrease is especially remarkable just below the 
dynamical transition temperature (or the mode-coupling temperature) 
\cite{Karmakar2009}. In our model, with the local dynamics, the relaxation 
time does not decrease with the system size even around the dynamical 
transition temperature $T_\mathrm d = 0.348(3)$, see \Fig{fig:s_relax_time}.
At lower temperature, the relaxation time rather increases with the system size.
The finite-size effect is more prominent in the relaxation time of 
the non-local dynamics we used for equilibrium computation, see \Fig{fig:s_relax_time}.
Intuitively, the finite-size effect is clearer because of the following reason. 
The relaxation process at moderately low temperature is mixed with many modes,
with each contribution. The slowest mode of the relaxation process, which 
diverges towards the transition temperature and has strong finite size effects, is buried 
in many other modes due to its small contribution in the dynamics of inefficient algorithms, 
including simple local dynamics. The relaxation time shows a finite-size effect only when the 
slowest mode contribution becomes very large, which is very low temperature. 
Efficient algorithms, on the other hand, can reduce contributions of modes faster than 
the slowest mode but still very slow, and thus a finite-size effect that the slowest mode has
is seen more clearly. This is actually quite general in other models, even in simple ferromagnetic
spin models, by comparing the dynamics of the simple Metropolis algorithm and more
efficient algorithms such as the over-relaxation, for instance.

\begin{figure}[htbp]
\center \includegraphics[width=.5\linewidth]{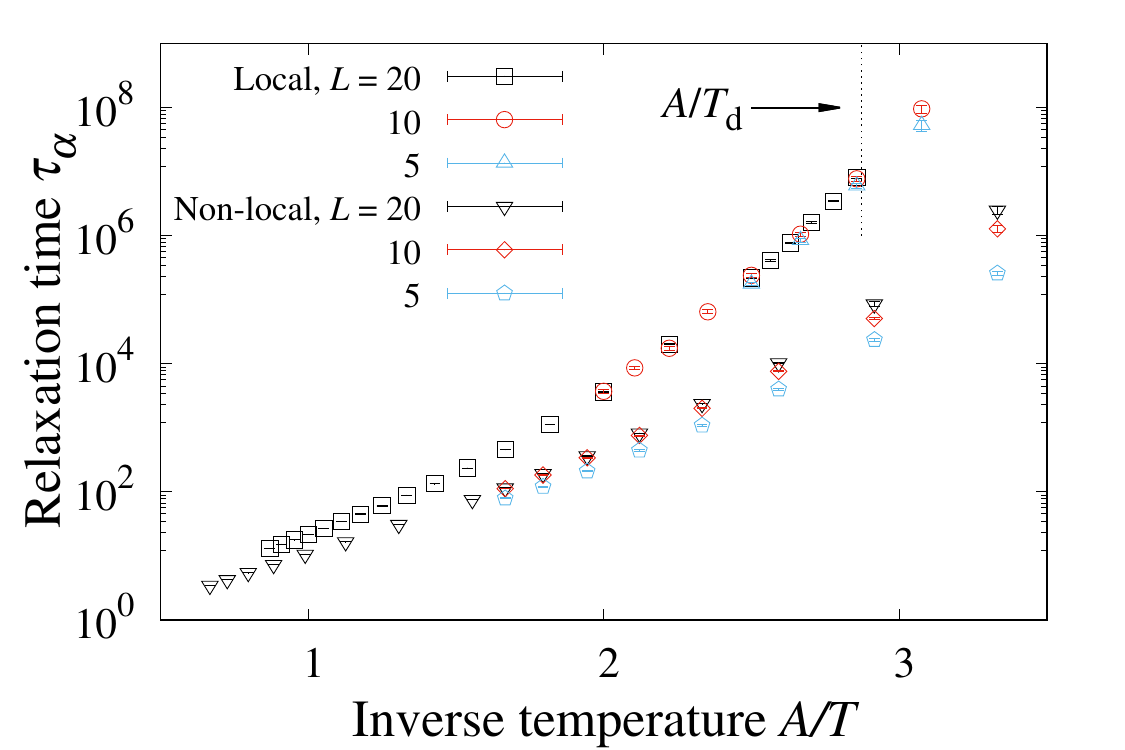}%
\caption{
The relaxation time as a function of inverse temperature for $L = 5, 10$, and $20$
for the local and non-local dynamics. The inverse dynamical transition temperature 
$A/T_\mathrm d$ is estimated in Supplemental Item 2. 
}
\label{fig:s_relax_time}
\end{figure}

\fi

\end{document}